\documentclass{aa}  

\usepackage[varg]{txfonts}
\usepackage{lineno}
\usepackage{color}
\usepackage{amssymb}
\usepackage{txfonts}
\usepackage{lscape}
\usepackage{float}
\usepackage{amsmath}
\usepackage{color}
\usepackage{longtable}
\usepackage{journal_names}
\usepackage{rotating}
\usepackage{pdflscape}
\usepackage{dcolumn}
\usepackage{caption}
\usepackage{subcaption}
\usepackage{sidecap}
\usepackage[nameinlink]{cleveref}
\usepackage{xcolor}

\usepackage{color,soul}
\usepackage{amssymb}
\usepackage{lscape}
\usepackage{float}
\usepackage{enumitem}
\usepackage{mathtools}
\DeclareMathOperator{\MyProd}{\scalebox{1.4}{$\mathrm{I\kern-0.2ex I}$}}

\usepackage{subcaption}
\usepackage{caption}
\usepackage{xcolor}
\usepackage[export]{adjustbox}
\DeclareCaptionLabelFormat{continued}{Figure \ref{postage}}

\def\la{\mathrel{\hbox{\rlap{\hbox{\lower4pt\hbox{$\sim$}}}\hbox{$<$}}}}
\def\ga{\mathrel{\hbox{\rlap{\hbox{\lower4pt\hbox{$\sim$}}}\hbox{$>$}}}}

\def\arcmin{\hbox{$^\prime$}}
\def\arcsec{\hbox{$^{\prime\prime}$}}

\def\deg{{^\circ}}
\newcommand{\dg}{^{\circ}}

\newcommand{\MSUN}{${\rm M}_\odot$}

\newcommand{\kms}{{\,km\,s$^{-1}$}}

\newcommand{\HI}{\mbox{\normalsize H\thinspace\footnotesize I}}

\def\apjl   {{ApJL }}
\def\apjs   {{ApJS }}

\begin{document}

\nolinenumbers

\title{MeerKAT HI imaging of the jellyfish galaxy ESO\,137-001}

\author{M. Ramatsoku\inst{1,}\inst{2}\fnmsep\thanks{m.ramatsoku@ru.ac.za} \and  P. Serra\inst{2} \and M. Sun\inst{3} \and O.M. Smirnov\inst{1,}\inst{4,}\inst{5} \and S. Makhathini\inst{6}}

\institute{Centre for Radio Astronomy Techniques and Technologies (RATT), Department of Physics and Electronics, Rhodes University, Makhanda, 6140, South Africa. 
         \and
             INAF- Osservatorio Astronomico di Cagliari, Via della Scienza 5, I-09047 Selargius (CA), Italy.
          \and 
             Department of Physics and Astronomy, University of Alabama in Huntsville, 301 Sparkman Drive, Huntsville, AL, 35899, USA.
          \and
             South African Radio Astronomy Observatory, Black River Park North, 2 Fir St, Cape Town, 7925, South Africa.
          \and 
             Institute for Radioastronomy, National Institute of Astrophysics (INAF IRA), Via Gobetti 101, 40129 Bologna, Italy.
           \and  
           School of Physics, University of the Witwatersrand, Johannesburg, Gauteng, South Africa.
           }

 \date{Received 02 October 2024 / Accepted 26 December 2024}


  \abstract{We present MeerKAT \HI\ observations of ESO\,137-001, a quintessential jellyfish galaxy with long multi-phase tails formed due to the interaction with the intra-cluster medium of its host galaxy cluster, ACO\,3627. Our observations reveal the presence of \HI\ in both the disc and outer regions of the galaxy for the first time, with a total \HI\ mass of ($3.5 \pm\ 0.4) \times 10^{8}$ \MSUN. ESO\,137-001 is at an advanced stage of gas stripping; it is extremely \HI\ deficient and seems to have lost 90\% of its initial \HI\ mass; about 2/3 of the surviving \HI\ is found at larger radius than expected for a normal \HI\ disc and forms $\sim40$\,kpc tail coincident with the tail detected at other wavelengths. Only $\sim10$\% of the surviving \HI\ is still found within the stellar disc, consistent with the expectation of an outside-in truncation due to ram pressure. Similarly to other jellyfish galaxies, ESO137-001 has a high star formation rate for the low amount of \HI\ detected. We measure an \HI\ depletion time of 0.29 Gyr. However, when taking into account the total gas (\HI\ + H$_2$) content, the depletion time is consistent with typical values measured in nearby spiral galaxies. This suggests that ESO\,137-001 is at its current stage of ram pressure interaction characterised by an efficient \HI\ stripping, rather than an enhanced conversion of \HI\ to H$_2$, which was recently observed in some other jellyfish galaxies.}


\maketitle
%

\section{Introduction}

One of the main objectives of ongoing research into galaxy evolution is to understand the origins and operational mechanisms that lead to the formation of non-star forming early-type galaxies. Higher density galaxy environments such as groups and clusters play a crucial role in this process and are essential to these investigations (\citealp{Dressler1980, Postman1984, Peng2010, Whitaker2012, Foltz2018}). Examples of galaxies transforming due to interactions with their host cluster environment have been reported; these are known as jellyfish galaxies (\citealp{Yagi2007, Chung2009, RSmith2010, Ebeling2014, Fumagalli2014, Poggianti2017nat}). They form as a result of the hydrodynamical interaction between the intracluster medium (ICM) and the interstellar medium (ISM) as they fall into the cluster centre (e.g. \citealp{Gunn1972}). Numerical wind-tunnel simulations have demonstrated that the formation of jellyfish galaxies results from a fast-moving galaxy in a dense ICM. The ram-pressure generated causes the ISM confined by the gravitational potential well of the galaxy to be pushed out of the stellar disc, resulting in highly asymmetric morphologies with extended stripped tails of debris material (\citealp{Roediger2008, Kapferer2009, Ramos2018}). The stripped tails have been observed and reported in various wavelengths, including X-ray (\citealp{Vikhlinin2005, Sun2010, Poggianti2019xray}), UV and optical wavelengths (\citealp{Cramer2019, Yagi2010, Yoshida2002}), radio continuum (\citealp{Gavazzi1995, Crowl2005, Chen2020, Roberts2021}), as well as in ionised gas (\citealp{Yagi2007, Fossati2012, Poggianti2017, Fossati2018, Cramer2019, Fumagalli2014, Luo2023}), molecular gas (CO; \citealp{Jachym2014, Verdugo2015, Jachym2017, Zabel2019, Moretti2020, Moretti2023}), and neutral atomic gas (\HI; \citealp{Gavazzi1989, Dickey1997, Kenney2004, Chung2007, Serra2013, Ramatsoku2019, Ramatsoku2020, Deb2020, Loni2021}). 

Many jellyfish galaxies have recently been discovered by detecting their ionised gas. This was achieved through extensive surveys, including the Virgo Environmental Survey Tracing Ionised Gas Emission (VESTIGE; \citealp{Boselli2018}) conducted with the MegaCam instrument on the Canada-France-Hawaii Telescope and the GAs Stripping Phenomena (GASP; \citealp{Poggianti2017}) carried out with the Multi Unit Spectroscopic Explorer (MUSE) Integral Field spectrograph on the Very Large Telescope. These galaxies have also been extensively investigated through the LOFAR Two-Meter Sky Survey (LoTSS) in radio continuum (\citealp{Roberts2021}). Attempts have also been made to understand the properties of jellyfish using the TNG50 gravity+magnetohydrodynamical simulations (\citealp{Pillepich2019, Nelson2019}). From these observations and simulations, it has been demonstrated that jellyfish galaxies continue to form stars during the ram-pressure stripping processes, which removes their gas reservoir. Star formation takes place in not only the expected location of galaxy discs but also the stripped tails. Studies have also shown that the ram pressure of most galaxy clusters is generally not strong enough to strip the dense star-forming molecular gas, at least not to the extent that would result in the formation of long tails (\citealp{Lee2017, Cramer2020}). Instead, the stars in their tails are thought to form in situ from the stripped \HI\ gas that has cooled to a cold and dense phase (\citealp{Moretti2020}). The question of how star formation can occur given the diffused nature of stripped gas has been explored through studies demonstrating that ordered magnetic fields play a role in facilitating this process (e.g. \citealp{Ettori2000, Asai2005, Eckert2017}). This was demonstrated for a jellyfish, JO206, in the IIZW108 cluster by \citet{Muller2021} wherein it was reported that magnetic fields prevent momentum exchange and heating, thus shielding the neutral gas and allowing it to cool. The findings of \citet{Muller2021} are partially supported by studies that have reported an efficient conversion of cold and diffuse \HI\ to dense CO gas in jellyfish galaxies (\citealp{Jachym2014, Jachym2019, Moretti2020, Moretti2023}). However, the complex interplay between gas heating, compression, cooling, and star formation in tails remains poorly understood.

The efficiency of star formation activity in these gas-stripped galaxies remains a topic of ongoing debate.
For instance, \HI\ observations conducted on two confirmed jellyfish galaxies detected at optical wavelengths, namely, JO201 and JO206, revealed a coincidence between the spatial distribution and kinematics of H$\alpha$ and \HI, as well as enhanced global and resolved star formation activity at kpc-scales in the disc and tail. Simulations which account for galactic winds and cooling found that ram-pressure stripping enhanced the overall SFR in the compressed disc regions and the stripped tail \citep{Kronberger2008, Kapferer2009}. However, the enhanced SFR is not found in all cases; \citet{Deb2020, Deb2022} found that in other optically selected jellyfish galaxies, JO204 and JW100, the spatial distribution of \HI\ and H$\alpha$ was distinct, with significant kinematic decoupling observed. Moreover, galaxies undergoing ram-pressure stripping in the Virgo cluster do not have an enhanced star formation rate (\citealp{Yoon2017}, \citealp{Brown2023}). In a related study, \citet{Mun2021} used the WISE 22$\mu$ m flux densities to quantify star formation activity in ram-pressure stripped galaxies, revealing that it is suppressed compared to non-ram-pressure stripped galaxies. Other simulations, including the gravity+magnetohydrodynamical simulations conducted by \citet{Goller2023} on nearly a thousand simulated jellyfish galaxies, have also failed to provide evidence of enhanced SFR for these galaxies.

The differing findings from various studies highlight the lack of a comprehensive understanding of the influence of ram pressure on jellyfish star formation activity. Therefore, reliable measurements of the star formation activity in the bodies and tails remain crucial. Since \HI\ is highly susceptible to ram-pressure stripping and serves as the primary component in the formation of stars, it remains a valuable method for assessing the efficiency of star formation in jellyfish galaxies. However, the limited availability of \HI\ data for these galaxies highlights the need for more in-depth, individual studies to improve our understanding of this process.

~
\\
ESO\,137-001 is one of the most striking jellyfish galaxy undergoing ram-pressure stripping due to its host environment. It is located in the Norma cluster (ACO\,3627; \citealp{Abell1989}) within the Great Attractor region \citep{Dressler1987GA}. ACO\,3627 is a massive structure with a dynamic mass of  M$_{\rm dyn} \sim10^{15}$ \MSUN\ at a redshift of $z$ = 0.01625 \citep{Woudt2008}. ESO\,137-001 lies near the centre of ACO 3627, at a projected distance of 14.5\arcmin\ ($\sim$200 kpc) from the central BCG galaxy, ESO\,137-006 \citep{Sun2007}. Its line-of-sight velocity is $v = 4647$ \kms\, which is a 200 \kms\ difference from the cluster centre, indicating that it is moving along the plane of the sky \citep{Luo2023}.\\

This jellyfish galaxy has been extensively observed at various wavelengths. Its initial discovery as a jellyfish was made through X-ray observations conducted with XMM and Chandra \citep{Sun2006}. These observations revealed a long, bifurcated X-ray tail extending from the galaxy disc, with the northern component extending $\sim$80 kpc and a shorter southern component. Further observations using the Southern Astrophysical Research (SOAR) telescope revealed an additional 40 kpc H$\alpha$ tail extending from the galaxy disc and pointing away from the cluster centre \citep{Sun2007}. This H$\alpha$ tail was also imaged using the Multi Unit Spectroscopic Explorer (MUSE) on the UT4 Very Large Telescope by \cite{Fumagalli2014} and recently by \citet{Luo2023}, who demonstrated that the tail extends much further than previously reported. Its spatial distribution also coincides with the X-ray tail. Within this stripped H$\alpha$ tail, compact, bright knots with the characteristic line ratios, densities, temperatures and metallicity consistent with gas photoionised by young stars were identified (\citealp{Sun2010, Fumagalli2014, Fossati2016, Luo2023}). The investigation of the star-formation properties of ESO\,137-001 using the Infrared Spectrograph (IRS) on the Spitzer Space Telescope revealed a warm H$_{\rm 2}$ 20\,kpc tail with a mass of $\sim$10$^{7}$\MSUN\ that is aligned with both the X-ray and H$\alpha$ tails \citep{Sivanandam2010}. Follow-up observations with the single dish Atacama Pathfinder Experiment (APEX) revealed CO(2-1) emission in three positions along the 40\,kpc H$\alpha$ tail and detected a total H$_2$ mass of $\sim10^9$\,\MSUN\ indicating that there is abundant cold, dense gas forming stars in both the disc and tail \citep{Jachym2014}. The Atacama Large Millimeter Array (ALMA) provided high-resolution images of the cold molecular gas traced by CO(2-1) emission in ESO\,137-001, revealing a significant amount of this gas phase in both the disc and tail of the galaxy. Its spatial distribution in the tail exhibited clumpy CO overdensities, which likely formed in situ and partly overlapped with H\textsc{ii} regions \citep{Jachym2019}. A detailed study of the star-forming regions was undertaken with Wide Field Camera 3 (WFC3) and the Advanced Camera for Surveys (ACS) on the \textit{Hubble} Space Telescope (HST) in the F275W, F475W, F814W and F160W filters. The data showed a correlation between the H\textsc{ii} regions and blue stellar clumps within 0.2\,kpc, with the stellar regions measuring a mass of $\sim 10^{4}$ \MSUN\ and ages less than 100 Myr \citep{Waldron2023}.

The amount of CO, ionised gas detected, and young stellar ages in the tail suggest the presence \HI. Observations of the \HI\ emission of the host cluster ACO\,3627 were conducted with the Australian Telescope Compact Array (ATCA; \citealp{Frater1992}). Observations were taken with three 30\arcmin\ (FWHM of the ATCA primary beam) pointings within one Abell radius of ACO\,3627 (1.75$\deg$; \citealp{Woudt2008}) with one of the pointing containing ESO\,137-001 centred at the location of the bright central galaxy (BCG) ESO\,137-006. They were sensitive to rms noise levels of $\sim$1\,mJy/beam at a resolution of 15\arcsec. However, no \HI\ was detected in ESO\,137-001, resulting in an upper limit on the total \HI\ mass of $\sim 6.0 \times 10^{8}$\,\MSUN\ at the 3$\sigma$ level for a 150\kms\ linewidth \citep{Vollmer2001}. 

~
\\
In this paper, we improve the status of the \HI\ data available for ESO\,137-001. The outcomes of our \HI\ observations with MeerKAT (\citealp{Jonas2016, Mauch2020}) are presented. We quantify the amount of \HI\ present in ESO\,137-001 and examine the spatial distribution of the extraplanar gas in the ram-pressure stripped tail of the galaxy and in the disc. Furthermore, we use the acquired \HI\ data to investigate the correlation between the cold, warm, and hot gas phases and assess the efficiency of star formation under ram-pressure stripping conditions.

The structure of the paper is as follows: In Sect.\,\ref{observations}, we present the \HI\ observations and describe the data reduction procedure. The details of the \HI\ source finding procedure and measurements of the total \HI\ content are given in Sect.\,\ref{sourcedets}. We compare the \HI\ data to the available multi-wavelength data in Sect.\,\ref{multiwav_comp}. In Sect.\ref{sfr_gas}, we discuss the star formation activity of ESO\,137-001 in relation to its gas content and summarise the main results in Sect.\,\ref{summary}. 

~
\\
Throughout this paper we adopt a \citet{Chabrier2003} initial mass function (IMF) and assume a $\Lambda$ cold dark matter cosmology with $\Omega_{\rm M} = 0.3, \Lambda_{\Omega} = 0.7$ and a Hubble constant, H$_{0}$ = 70 \kms\ Mpc$^{-1}$ .

\section{H\textsc{i} observations and data reduction}\label{observations}
ACO\,3627 was observed with MeerKAT \citep{Jonas2016, Mauch2020} in May 2019 (project ID SCI-20190418-SM-01) with 64 antennas using the L-band; 856 -- 1712 MHz, for a total of $2 \times\ 7$ hours on-source. We used the 4k mode of the SKARAB correlator, which was the only correlator available at the time. It samples the observed band with 4096 channels that are 209-kHz-wide in full polarisation. This resolution was adequate to analyse the spatial distribution of detected \HI. Observations were centred on the bright central galaxy (ESO\,137-006; \citealp{Ramatsoku2020}) at ($\alpha_{\rm J2000}, \delta_{ \rm J2000}$) = (16h15m11s, -60d54m21s), which is about 14.5\arcmin\ from ESO\,137-001. This observation scheme was used to mitigate the impact of direction-dependent calibration errors associated with this bright radio source (167 mJy at 1398 MHz; see \citealp{Ramatsoku2020}). To calibrate the bandpass, complex gains and absolute flux-scale of the instrument, we observed PKS 1934-638 (Jy) for 7\,mins after every 60 mins on-source pointings for a total of 1.75 hours. 

We processed the data over the frequency range covering the Norma cluster and its background, 1356.620 - 1439.642 MHz, centred at 1398.055 MHz. 
The \emph{uv}-data were reduced using the Containerised  Automated Radio Astronomical Calibration (\textsc{CARACal}; \citealp{Josza2022}) pipeline\footnote{https://caracal.readthedocs.io}. This data reduction pipeline is built using the radio interferometry framework, \textsc{stimela}\footnote{https://github.com/SpheMakh/Stimela} \citep{makhathini2018}, which is based on container technologies and Python. The framework supports the combining of various open-source radio interferometry software packages in the same script. The calibrator and target visibilities were flagged with \textsc{AOFlagger} \citep{Offringa2010} based on the stokes Q visibilities.  We used \textsc{casa} \citep{McMullin2007} tasks \textit{bandpass} and \textit{gaincal} to determine the antenna-based time-independent complex flux, bandpass and frequency-independent gains, and \textit{fluxscale} was used to bootstrap flux scale and gain amplitudes. The resulting complex gain and bandpass solutions were applied on-the-fly to the target visibilities with the \textsc{casa} task \textit{applycal} during the splitting of the target visibilities with \textit{mstransform}. ESO\,137-001 exhibits a strong, extended continuum emission (see; \citealp{Koribalski2024}). Furthermore, its proximity to a bright radio galaxy with diffuse, extended lobes necessitated careful continuum subtraction. This step was critical to prevent residual continuum artefacts that could be misinterpreted as faint \HI\ emission within the galaxy \citep{Grobler2014}.
To ensure accurate continuum subtraction, we created a model of the continuum sources by imaging and self-calibrating the continuum emission of the target iteratively. The imaging was done with \textsc{WSclean} in Stokes\,I using multi-scale cleaning (\citealp{Offringa2014}, \citealp{offringa2017}) with the Briggs \textit{robust} weighting parameter set to $r$ = 0 and cleaning down to 0.5$\sigma$ within a clean mask made with \textsc{sofia} \citep{Serra2015, Westmeier2021}. We self-calibrated the gain phase with \textsc{cubical} \citep{Kenyon2018} with a solution interval of 128 seconds. The resulting final continuum model was subtracted from the calibrated visibilities with the \textsc{msutils} package. We then subtracted residual continuum emission by fitting a 3rd-order polynomial to the visibilities. This high-order polynomial fit was necessary to sufficiently subtract all the continuum emission. The continuum subtracted \emph{uv}-data was then Fourier transformed into a final \HI\ cube with a pixel size of 2\arcsec\ and a field of view of 1 deg$^{2}$. We used a weighting with \textit{Briggs} robust parameter, $r$ = 0.5 and 10\arcsec\ \emph{uv}-tapering to optimise the surface brightness sensitivity. The sidelobes of the synthesised beam were removed by iteratively using \textsc{SoFiA} to produce 3D clean masks and imaging with \textsc{WSclean} while cleaning within the masks down to 0.5$\sigma$. The resulting \HI\ cube has a restoring Gaussian PSF with FWHM of $22.3\arcsec\ \times\ 16.8\arcsec$ and a position angle of PA = $150\dg$ with an rms of $\sigma =$ 0.06 mJy/beam. With these data, we reach a column density sensitivity of N$_{\rm HI} = 2.3 \times 10^{19}$ atoms cm$^{-2}$ at 3$\sigma$ at a location of ESO\,137-001 in the pointing field assuming a linewidth of 45\kms\ \citep{Meyer2017}. A summary of the \HI\ observation and cube properties is listed in Table.\,\ref{hiobs_summary}

\begin{table}
\caption {Summary of the H\textsc{i} observations.}\label{hiobs_summary}
\begin{tabular}{llllll}
\hline
 Property        & ESO\,137-001   \\  
\hline\hline
Date             &  08 and 09 May 2019    \\
Project ID.      & SCI-20190418-SM-01\\
~~~~$\alpha$ (J2000) &$16^{\rm h}15^{\rm m}03^{\rm s}.8$ \\
 ~~~~$\Delta$(J2000)&  $-60\dg54\arcmin26\arcsec.0$   \\

Processed freq. range & (1356 - 1439) MHz&         \\
Calibrators: & \\
~~~~Gain, Flux, bandpass& J1939-6342               \\
On-source integration & 2 $\times$ 7 hrs \\ 
\hline
\HI\ cube properties\\
\hline
Channel width & 44.5\,\kms at $z=0$ \\
Cube weighting & robust 0.5, taper = 10\arcsec\\
Sensitivity (r.m.s per channel) & 0.06 mJy beam$^{-1}$\\
Restoring Beam (FWHM) (P.A) & $22.3\arcsec \times 16.8\arcsec\ $ ($150\dg$) \\ 
\hline\hline\\
\end{tabular}
\end{table}

\section{The H\textsc{i} gas emission of ESO\,137-001}\label{sourcedets}
\begin{figure*}
   \centering 
   \includegraphics[width=0.70\textwidth]{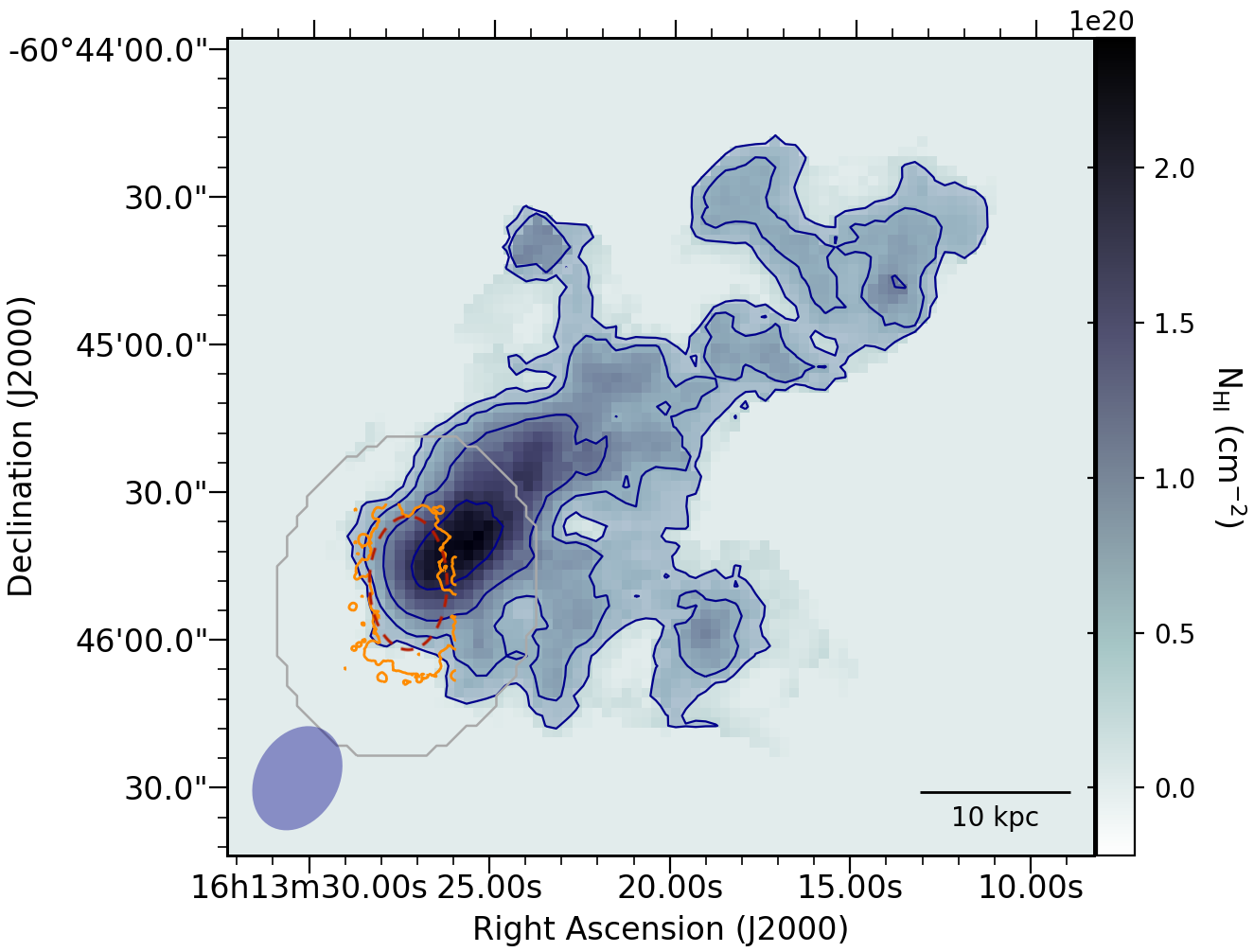}
 \caption{Integrated H\textsc{i} map of ESO\,137-001. The blue contours are at H\textsc{i} column densities of 2.3, 4.6, 9.2, 18.4 $\times$ 10$^{19}$ atoms/cm$^{2}$ and the blue ellipse denotes the FWHM beam size of $22\arcsec \times 16\arcsec$ (PA = 150$\dg$). The orange contour represents the galaxy's optical disc at the 22 mag arcsec$^{2}$ isophote in the $r$-band \citep{Fossati2016}. The grey contour outlines a model \HI\ disc defined at the sensitivity limit of our observations, N$_{\rm HI}$ =2.3 $\times$ 10$^{19}$ atoms/cm$^{2}$ and the red ellipse delineates the H\textsc{i} stripping radius (see Sect.~\ref{sec:HIgasdist}).}\label{HIdist}
 \end{figure*}

Our search for the \HI\ emission of ESO\,137-001 in the final \HI\ cube was conducted with \textsc{sofia} using the smooth and clip (S+C) method. We employed Gaussian spatial kernels, 4\arcsec, 12\arcsec, and 24\arcsec, as well as velocity boxcar smoothing kernels of 44 km/s, 132 km/s, and 220 km/s. We set the threshold for accepted \HI\ emission to be brighter than 3$\sigma$ with a reliability of 0.90. The latter is computed by \textsc{sofia} by comparing the distribution of negative and positive sources to distinguish real emission from noise peaks. The first-ever detection of \HI\ in ESO\,137-001 in Fig.\,\ref{HIdist} shows the extracted \HI\ line emission column density.

We measured a total integrated flux of (0.30 $\pm$ 0.03) Jy \kms\ from the \HI\ map shown in Fig.\,\ref{HIdist}. Adopting the luminosity distance reported in \citet{Jachym2014} of D = 69.6 Mpc for ESO\,137-001 result in a total \HI\ mass, M$_{\rm HI} =$ ($3.5 \pm\ 0.4) \times 10^{8}$ \MSUN\ \citep{Meyer2017}. The uncertainty in the \HI\ mass was calculated through error propagation accounting for the noise of the cube, the number of independent cube voxels included in the \HI\ detection mask, and a 10 per cent flux calibration error \citep{Serra2023}. The measured mass is consistent with the \HI\ mass upper limits from the previous ATCA observations \citep{Vollmer2001}.

\subsection{H\textsc{i} deficiency}\label{hidef}
\begin{figure}
   \centering
 \includegraphics[width=\columnwidth]{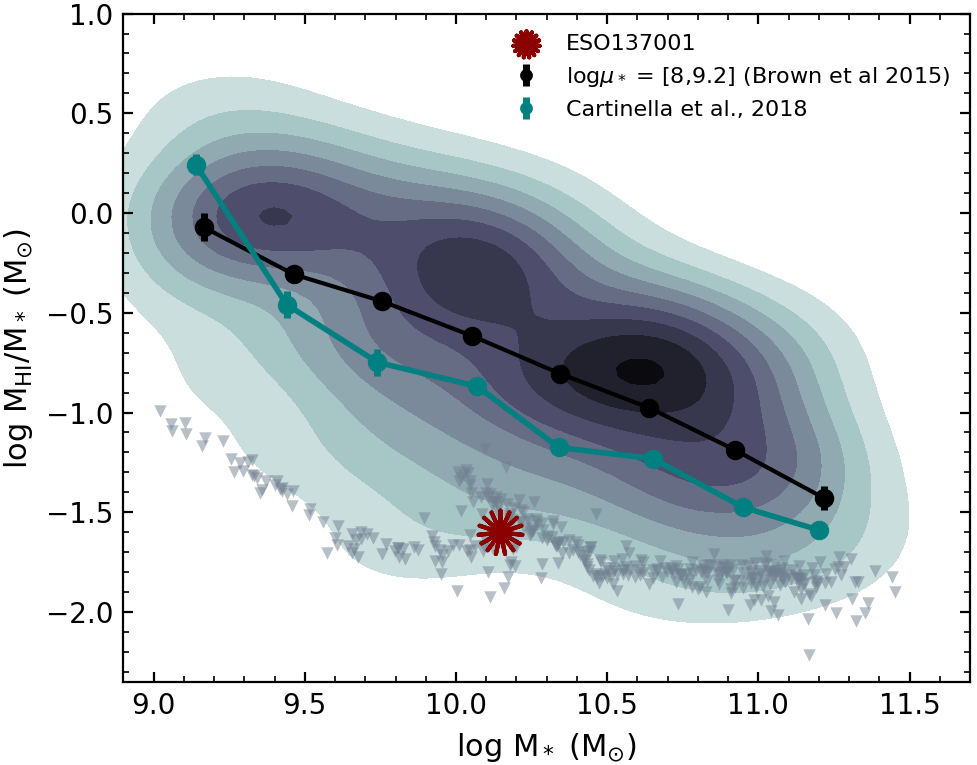}
    \caption{H\textsc{i} fraction as a function of the stellar mass. Field galaxies from xGASS are shown in grey where the contours are \HI\ detected galaxies, and the triangles represent the non-detections. The xGASS scaling relations based on average gas fraction per stellar mass bin are shown in cyan, while the black points represent galaxies with stellar surface brightness comparable to that of ESO\,137-001. The scatter in the mass bins is illustrated by the error bars. ESO\,137-001 is represented by the red asterisk.}\label{HIstack}
 \end{figure}

We measured the total amount of \HI\ stripped by ram-pressure by comparing the observed \HI\ mass with the expected, namely, the \HI\ deficiency; Def$_{\rm HI}$\ defined by \citet{Haynes1984} as;
\begin{equation}
\mathrm{Def_{\rm HI} = log(M_{HI,exp}) - log(M_{HI,obs})}   
\end{equation}
The expected \HI\ mass was determined using scaling relations based on magnitudes and diameters at optical and near-infrared (NIR) wavelengths, as expressed in equations 5 and 6 of \citep{Denes2014}. Based on this prescription, galaxies are typically classified as HI-deficient when their Def$_{\rm HI}$ value exceeds 0.6, roughly four times less \HI\ than the average.
ESO\,137-001 is situated within the Zone of Avoidance at low galactic latitudes, where optical wavelengths suffer dust extinction. To circumvent this issue, we used the galaxy $K-$band magnitudes and diameter data obtained from the 2MASS Extended Source Catalog (2MASX; \citealp{Jarrett2000}). This wavelength is less impacted by dust and has an upper extinction limit of A$_{K} = 1.5$ mag at the low galactic latitude of ESO\,137-001. The galaxy has an absolute magnitude, M$_{\rm K}$ = -21.97 mag \citep{Sun2007} and diameter, d$_{K}$ = 5.5 arcsec \citep{Skrutskie2006}. Using these and the corresponding fit parameters in Table\,2 of \citet{Denes2014}, we measure an expected mass of $ 4.6 \times 10^{9}$ \MSUN\ and $5.6 \times 10^{9}$\,\MSUN\ from the respective magnitudes and diameter based scaling relations. These correspond to Def$_{\rm HI}$ = 1.12 and 1.20, respectively. 
To validate these measurements, we also determined the \HI\ deficiency by comparing the observed \HI\ fraction (M$_{\rm HI}$/M$_{\ast}$) with the value expected based on the scaling relations derived for field galaxies in the extended GALEX Arecibo SDSS Survey (xGASS; \citealp{Catinella2018}). In Fig.\,\ref{HIstack}, we compare the observed \HI\ gas fraction with the stacked xGASS scaling relation for galaxies with a stellar surface density comparable to that of ESO\,137-001 (log$\mu^{*} = 8.6$ \MSUN/kpc$^{2}$) \citep{Brown2015}, and with the average \HI\ gas fraction scaling relation \citep{Catinella2018}. ESO\,137-001 lies 0.7 - 0.9 dex below the field population. Thus, the expected \HI\ mass of the galaxy is estimated to be up to M$_{\rm HI} \approx 2.8 \times 10^{9}$ \MSUN. 
These methods indicate a high degree of agreement regarding the fraction of \HI\ gas loss for this galaxy. Using these various techniques, we estimate that the galaxy has generally lost approximately 90\% of its initial \HI\ mass.

\subsection{The \HI\ gas distribution}\label{sec:HIgasdist}
As shown in Fig.\,\ref{HIdist}, the peak of the \HI\ surface brightness is offset by approximately 14\arcsec\ compared to the centre of the stellar disc. On the east side, the \HI\ distribution is truncated within the stellar disc. The \HI\ truncation is even more pronounced in the south of the stellar disc, potentially due to ram pressure having a northern component relative to the cluster centre. This is highly likely given the projection of the entire \HI\ distribution. On the west, \HI\ extends significantly beyond the stellar disc, forming a tail that stretches out to about 40 kpc from the galaxy centre. This \HI\ morphology is consistent with the well-established effects of ram-pressure stripping effects on ESO\,137-001 by the ICM in the Norma cluster.
The entire \HI\ tail is broad, measuring an average of $\sim15$ kpc in width in the north-south direction. However, at $\sim20$ kpc from the centre of the stellar disc, it becomes more pronounced, widening $\sim$30 kpc.
Further from the stellar disc, the \HI\ tail narrows to a width of approximately 5 kpc, forming a continuous structure. This may suggest that the galaxy has experienced ram pressure for an extended period, leading to the complete stripping of its outer, low-surface density gas disc. This scenario is plausible, as semi-analytic orbital models indicate the galaxy is roughly 100 Myr before its closest approach, moving at a velocity of 3000 \kms\ \citep{Jachym2014}.

\begin{figure}
   \centering
 \includegraphics[width=\columnwidth]{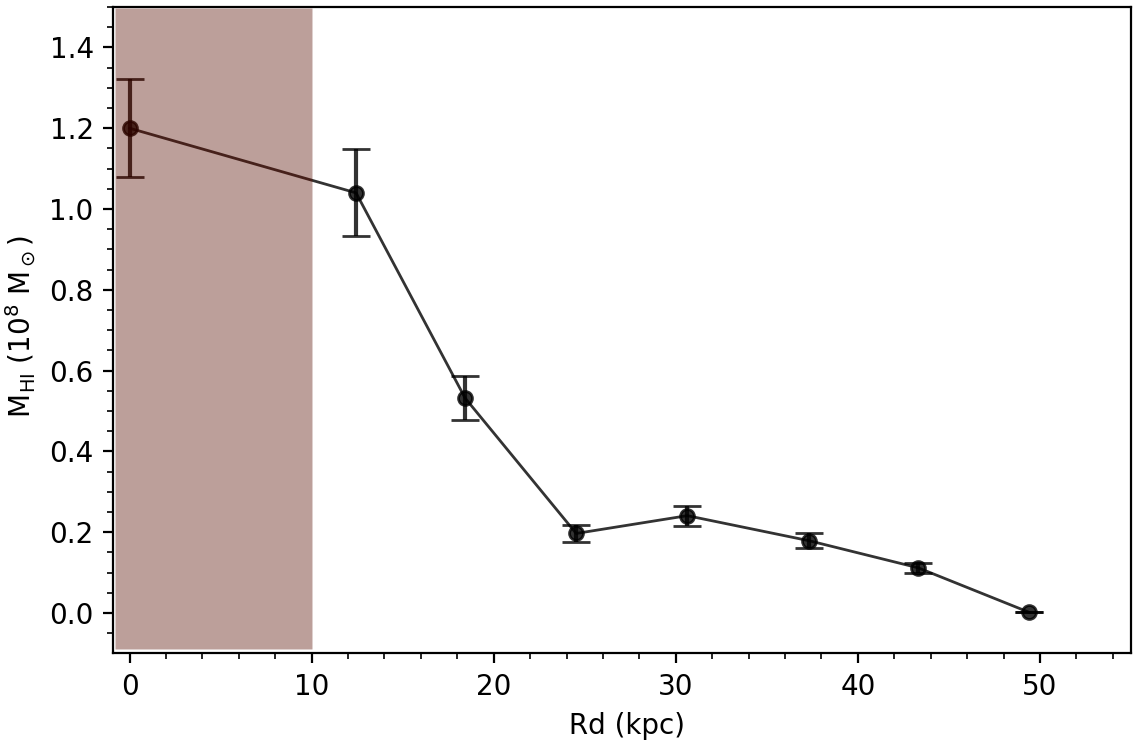}
    \caption{H\textsc{i} mass measured in 5 kpc bin widths along the H\textsc{i} tail as a function of the projected distance from the centre of the stellar disc of ESO\,137-001. The red shaded area indicates the mass within the H\textsc{i} disc.}\label{HImassbins}
 \end{figure}

In Sect.~\ref{hidef}, we found that ESO\,137-001 has lost up to $\sim$90\% of the initial mass. Fig.\,\ref{HIdist} shows that a significant fraction of its remaining \HI\ is not distributed on a settled disc. To measure how much \HI\ is found at larger distances than expected for a normal \HI\ disc of the same \HI\ mass, we applied the method from \citet{Ramatsoku2019}. We built a model assuming that all detected \HI\ is distributed on a settled disc and that ram pressure has not displaced any HI (i.e. no tail has formed). The model exploits the tight correlation between \HI\ mass and galaxy diameter, along with the predictable radial distribution of \HI\ (\citealp{Wang2016, Martinsson2016}). Based on the \HI\ size–mass relation parametrised by \citet{Wang2016}, 
\(\log\left(\frac{D_{\text{H I}}}{\text{kpc}}\right) = 0.51 \log\left(\frac{M_{\text{H I}}}{M_{\odot}}\right) - 3.32\), defined at a surface density of \(1 \, M_{\odot} \, \text{pc}^{-2}\). 
We used \HI\ surface density radial profile by \citet{Martinsson2016} as: \(\Sigma_{\text{H I}}(R) = \Sigma_{\text{H I}}^{\text{max}} e^{-\frac{(R - R_{\Sigma, \text{max}})^{2}}{2\sigma_{\Sigma}^{2}}}\),
where \(R_{\Sigma, \text{max}} = 0.2D_{\text{H I}}\) and \(\sigma_{\Sigma} = 0.18D_{\text{H I}}\) are fixed parameters. The only free parameter is \(\Sigma_{\text{H I}}^{\text{max}}\), which we set to \(0.4 \, M_{\odot} \, \text{pc}^{-2}\) to ensure that \(\Sigma_{\text{H I}}(D_{\text{H I}}/2) = 1 \, M_{\odot} \, \text{pc}^{-2}\). Using these parameters and assuming our observational conditions, we used the \textsc{3D-Barolo} package to model the settled \HI\ disc. For the inputs to the model, we assumed a typical \HI\ disc scale height of 0.3 kpc \citep{Randriamampandry2021}, and the position angle and inclination of the stellar disc as listed in Table~1 of \citet{Luo2023}. To ensure the model matched our observations, we convolved it with our point spread function (PSF) within \textsc{3D-Barolo}. The resulting radius of the model \HI\ disc with R$_{\rm mod, HI}$ = 10\,kpc (semi-major axis) was defined at the \HI\ column density sensitivity of our MeerKAT observations (N$_{\rm HI} = 2.3 \times 10^{19}$ atoms cm$^{-2}$) and is shown in Fig.~\ref{HIdist} (grey contour) where it is compared with the observed \HI\ flux density map. Any \HI\ emission in the observed map outside the grey contour is considered part of the \HI\ tail. We measured an \HI\ mass of $(2.3 \pm 0.2) \times 10^{8}$ \MSUN\ within this tail, which accounts for 66\% of the total \HI\ mass observed in this galaxy. As shown in Fig.~\ref{HImassbins}, about 50\% \HI\ in the tail is concentrated within a projected distance of 20 kpc from the stellar disc, where the tail is broader. In this region, we measure an \HI\ mass of \(M_{\text{HI}} \approx 1.1 \times 10^{8} \, \text{M}_{\odot}\). The rest is spread out beyond this distance, where the tail narrows.

To assess whether the \HI\ morphology is consistent with ram-pressure stripping effects, we identified the approximate radius outside which \HI\ gas has been removed from the stellar disc. This radius, shown by the red ellipse in Fig.~\ref{HIdist}, is $\sim3.6$ kpc (note that this is the observed radius, not deconvolved with the HI resolution; we take care of the effect of resolution as discussed below). Inside this region, we observe an \HI\ mass of $(2.7 \pm 0.2) \times 10^{7}$ \MSUN, or about 8\% of the total observed \HI\ mass in ESO\,137-001.
Using the modelling approach outlined previously; we assumed an initial (pre-stripping) \HI\ mass of $2.0 \times 10^{9}$ \MSUN\  (approximated from scaling relation in Sect.\ref{hidef}) and measured an extent with a radius of 20.6\, kpc for the model disc, defined at our \HI\ column density sensitivity as above. We then truncated this model disc at various radii to simulate gas loss from ram-pressure stripping effects. By convolving each truncated model with the observational PSF, we adjusted the truncation radius until the model matched the red ellipse in Fig.\,\ref{HIdist}. The best match was found with a truncation radius of 1.2 kpc, corresponding to an observed radius of $\sim3.9$ kpc after convolution. 

Within the truncation radius, the model yielded an \HI\ mass of $1.1 \times 10^{7}$ \MSUN; this matches our observations within a factor of 2. Therefore, the fact that only about 10\% of the surviving \HI\ (and about 1\% of the initial, expected HI mass) is located within the stellar disc with an observed radius of $\sim$3.6 kpc is consistent with the expectation of outside-in truncation due to ram pressure.

\section{The H\textsc{i} and the multi-wavelength phases of ESO\,137-001}\label{multiwav_comp}
\begin{figure*}[htp!]
   \centering
   \includegraphics[width=\textwidth]{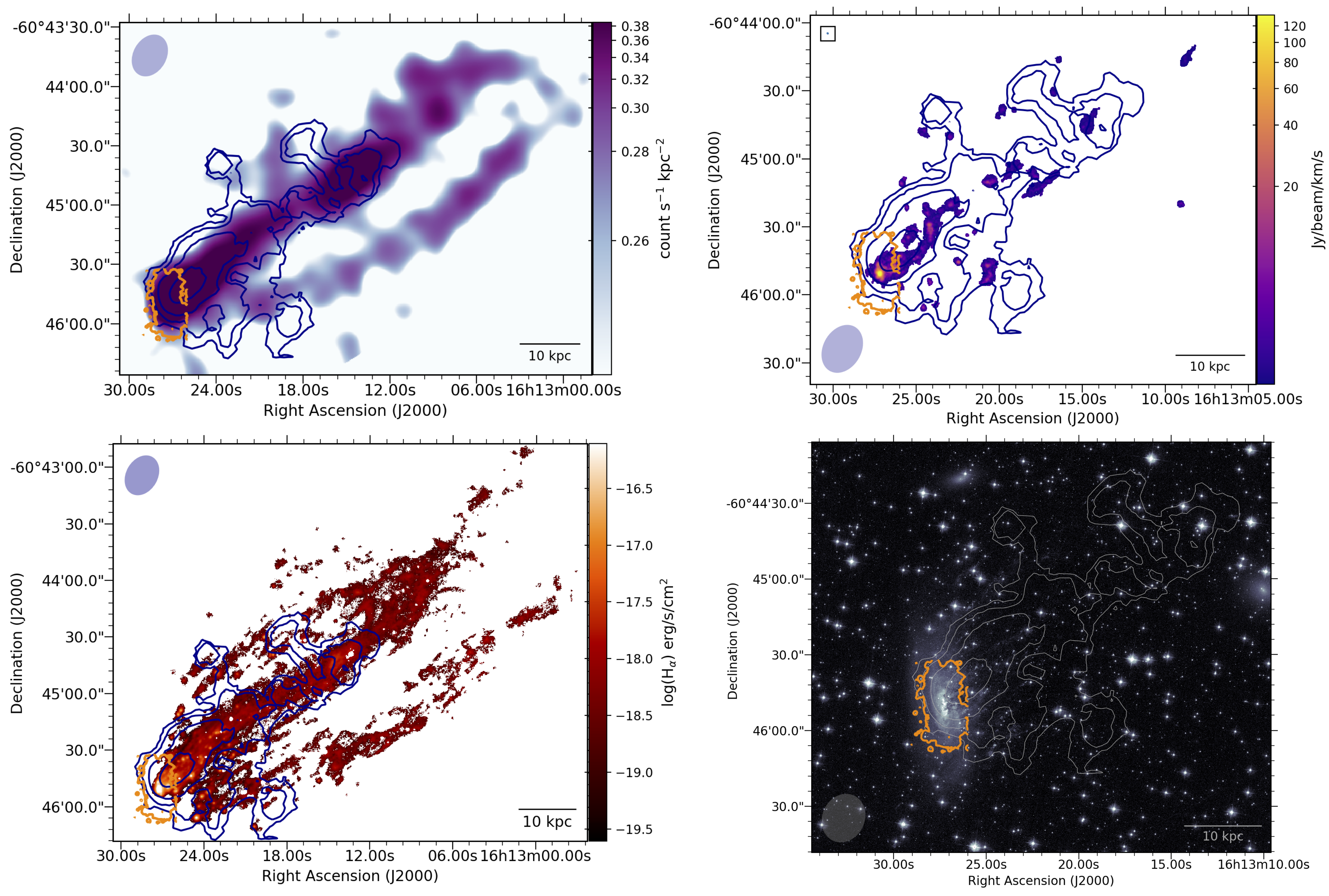}
    \caption{H\textsc{i} spatial distribution compared to X-ray, CO, H$\alpha$ and optical maps. The H\textsc{i} contours are at drawn at column densities of N$_{\rm HI}$ = 2.3, 4.6, 9.2, 18.4 $\times$ 10$^{19}$ atoms/cm$^{2}$. The orange contour is the galaxy disc, as in Fig\,\ref{HIdist}. \emph{Top left panel:} H\textsc{i} over Chandra 0.6–2.0 keV count X-ray image. \emph{Top right panel:} H\textsc{i} overlaid on the CO\,(2-1) flux intensity map from ALMA. \emph{Bottom left and right panels:} H\textsc{i} contours over the H$\alpha$ emission from MUSE and the HST WFC3 image (see also Fig.\,1 in \citealp{Waldron2023}), respectively.}\label{multiwav_fig}
 \end{figure*}

\begin{figure}[htp]
   \centering
 \includegraphics[width=70mm, height=95mm]{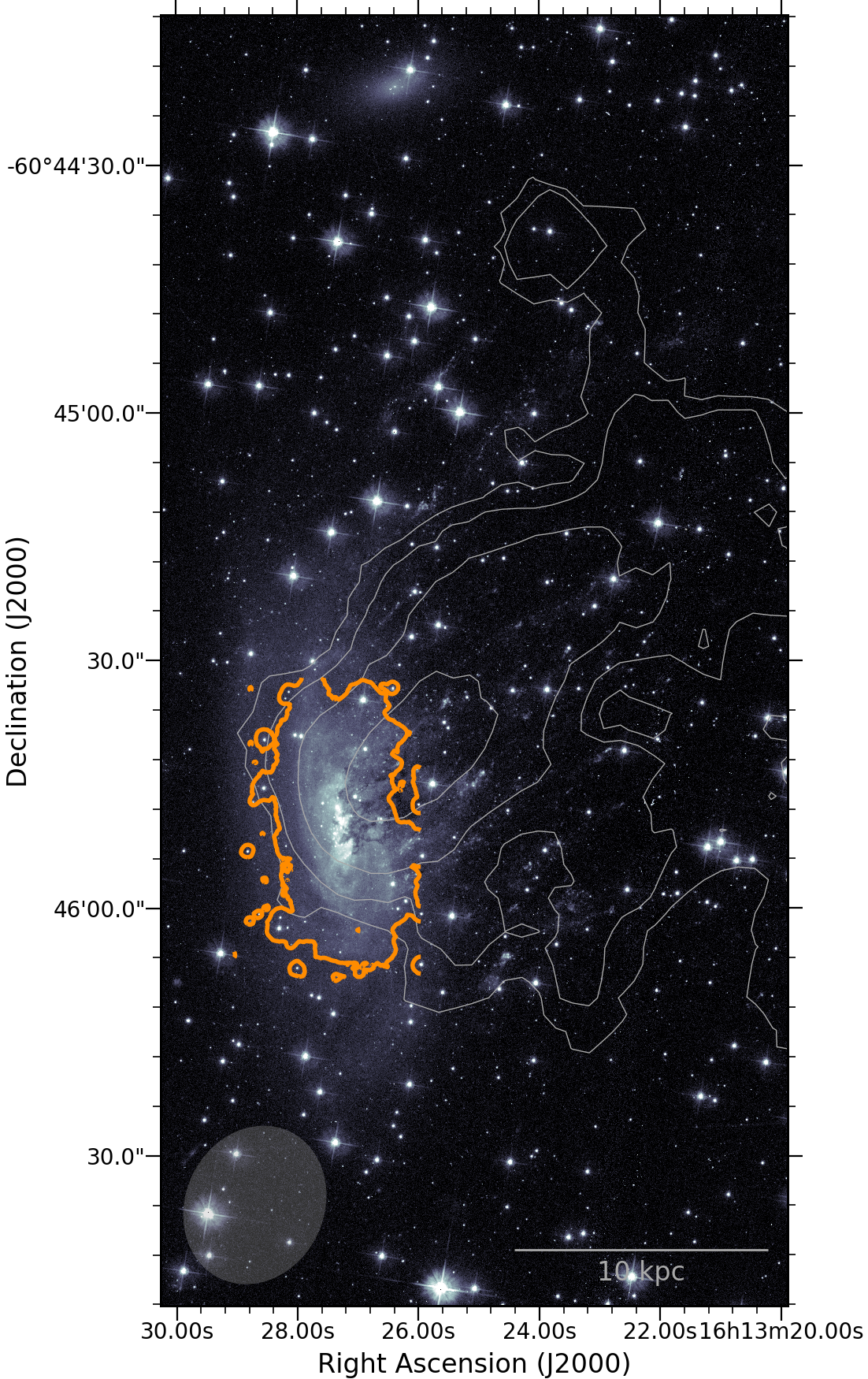}
    \caption{Zoomed-in image showing trails of young stars close to the galaxy disc with H\textsc{i} contours overlaid on the HST WFC3 image. The orange and grey contour levels are the same as in Fig.\,\ref{multiwav_fig}.}\label{zoomoptical}
 \end{figure}

In Fig.\,\ref{multiwav_fig}, we present a comparison of the \HI\ distribution in ESO\,137-001 with its X-ray, molecular, H$\alpha$ emissions, and star-forming regions (\citealp{Sun2010, Fumagalli2014, Jachym2019, Waldron2023}). 
The spatial resolution of \HI\ is generally lower than that of other gas phases, particularly CO and H$\alpha$, giving a false impression that the \HI\ distribution is more radially extended. Nonetheless, correlations can still be observed. 
All gas phases overlap almost entirely within and near the galaxy's stellar disc, where their surface brightness peaks. There is also a strong coincidence between all gas phases throughout the entire 40\, kpc long central component of the \HI\ tail. Beyond the \HI\ tail, there is  X-ray, CO, and H$\alpha$ emission. It is possible that the \HI\ in these farthest regions has been transformed into other gas phases or simply lost. This scenario is supported by the presence of a small clump of CO and ionised gas seen extending further than \HI. It could also be that the \HI\ is below our detection limit. 
The clumpy \HI\ protrusions extending perpendicular to the central tail show a local anti-correlation with the perpendicular components of the X-ray and CO emissions. However, X-ray emission is present between these protrusions, possibly tracing trails of stars formed in the galaxy (seen in the HST image), which may have consumed \HI\ in the region. This trend is less pronounced for H$\alpha$, where we observe hints of emission that appear to be associated with the \HI\ in these regions.
At optical wavelengths, bright young blue star regions with ages less than 10 Myr (see Fig.\,\ref{zoomoptical} and \citealp{Waldron2023}) are found within approximately 10 kpc of the stellar disc and almost overlap with \HI\ emission, particularly in the southern part of the tail. However, in the northern part, there are thin trails of stars that either do not coincide with \HI\ emission or lie just beyond the lowest \HI\ contours. These streams likely originate from molecular gas that formed in situ, as we also observe clumps of CO in those regions that are not linked to any detected \HI.
No F275W, F475W, and F814W HST band observations are available for comparison beyond 10\,kpc from the galaxy centre \citep{Waldron2023}. However, the overlap between \HI\ and H$\alpha$ suggests star formation activity beyond this distance from the disc.

\begin{figure*}
   \centering
 \includegraphics[width=180mm, height=123mm]{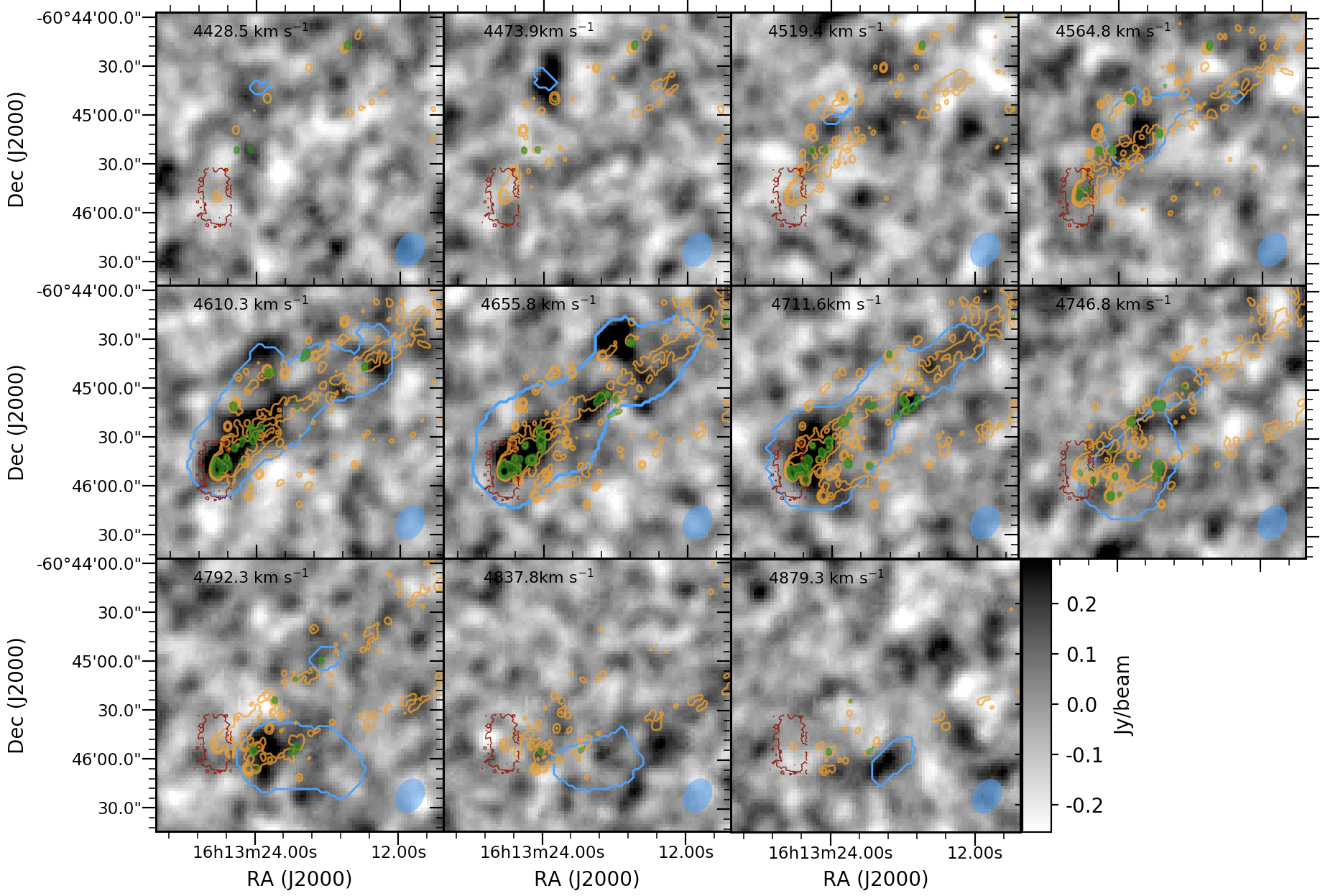}
 \caption{Channel maps of the H\textsc{i} cube of ESO\,137-001 compared with CO and H$\alpha$. The greyscale is linear and ranges from $-5\sigma$ to $5\sigma$ where $\sigma$ is the rms of the H\textsc{i} cube. Each channel is labelled with the channel velocity. The light blue contours represent the detection mask used to make the H\textsc{i} intensity (zeroth moment) maps. Green and orange contours are CO and H$\alpha$ emission at levels of  0.1, 0.3, 0.6 ... Jy beam$^{-1}$ and $1, 10, 100, ...\times 10^{-18}$ erg s$^{-1}$ cm$^{-2}$, respectively. The red contour is the stellar disc, as shown in Fig\,\ref{HIdist}.}.\label{Chmaps}
 \end{figure*}

Figure.\,\ref{Chmaps} shows the distribution of the \HI\ emission in velocity channels compared to the CO and H$\alpha$ emissions. We also show the \HI\ detection mask used to make the intensity map in Fig.\,\ref{HIdist}. Note that since the detection mask was generated from the smoothed cubes (see Sect.\,\ref{sourcedets}), some regions may outline emissions below the detection threshold in specific channel maps. To carry out this comparison, the CO cube and H$\alpha$ data were regridded to have the same velocity resolution as the \HI\ cube. However, due to the relatively low velocity resolution of the \HI\ cube ($\sim$44 \kms), we can only conduct a qualitative analysis of the general kinematics. Our analysis reveals a significant agreement in velocities between the three gas phases. The \HI\ disc and, more notably, the entire tail emission are seen at a velocity of $\sim$4655.8 \kms\ with the CO and H$\alpha$ showing the same velocity distribution. This redshift of all gas phases relative to the stars is consistent with the expected effect of ram pressure given that the galaxy is blueshifted relative to the ICM (the BCG lies at a velocity of 5444 \kms\ \citep{Saraf2023}.

We also find that within the optical disc, the H$\alpha$ emission is more extended in velocity by approximately 100 \kms\ than \HI\ and CO towards higher and lower velocities. The inner disc emission starts at velocities of $\sim$4470 to 4840 \kms. These regions typically exhibit clumpy H\textsc{ii} regions with velocity dispersions of around 30 \kms\ (\citealp{Fossati2016, Fumagalli2014, Luo2023}). Outside the optical disc, H$\alpha$ regions generally show higher velocity dispersions, with diffuse H$\alpha$ velocities largely consistent with the \HI.

\section{Gas content and star formation}\label{sfr_gas} 
The star formation rate of ESO\,137-001 has been determined from multiple wavelengths. \citet{Waldron2023} derived the total far-infrared luminosity of the galaxy from the \textit{Herschel} data and Galex NUV flux density, and estimated a total star formation rate of 0.97 \MSUN/yr. Additionally, from the WISE 22$\mu$ m flux density they estimated a total SFR of 1.39 \MSUN/yr. Therefore, the overall average SFR for ESO\,137-001 from these estimates is approximately 1.2 \MSUN/yr.\\
We evaluated the star formation rate of ESO\,137-001 based on its observed gas content in comparison to similar and previously studied jellyfish galaxies with \HI\ and CO data, namely JO201, JO206, and JW100 (\citealp{Ramatsoku2019, Ramatsoku2020, Deb2022}). Our reference field sample is extracted from xGASS for \HI\ and the extended CO Legacy Database for GASS (xCOLDGASS; \citealp{Saintonge2017}) for molecular gas. To ensure consistency with the jellyfish sample, we recalculated the H$_2$ mass from xCOLDGASS using a constant CO conversion factor, $\alpha_{\rm CO,MW}$ = 4.3 (K \kms\ pc$^2$)$^{-1}$.

\begin{figure}[h]
   \centering
 \includegraphics[width=\columnwidth]{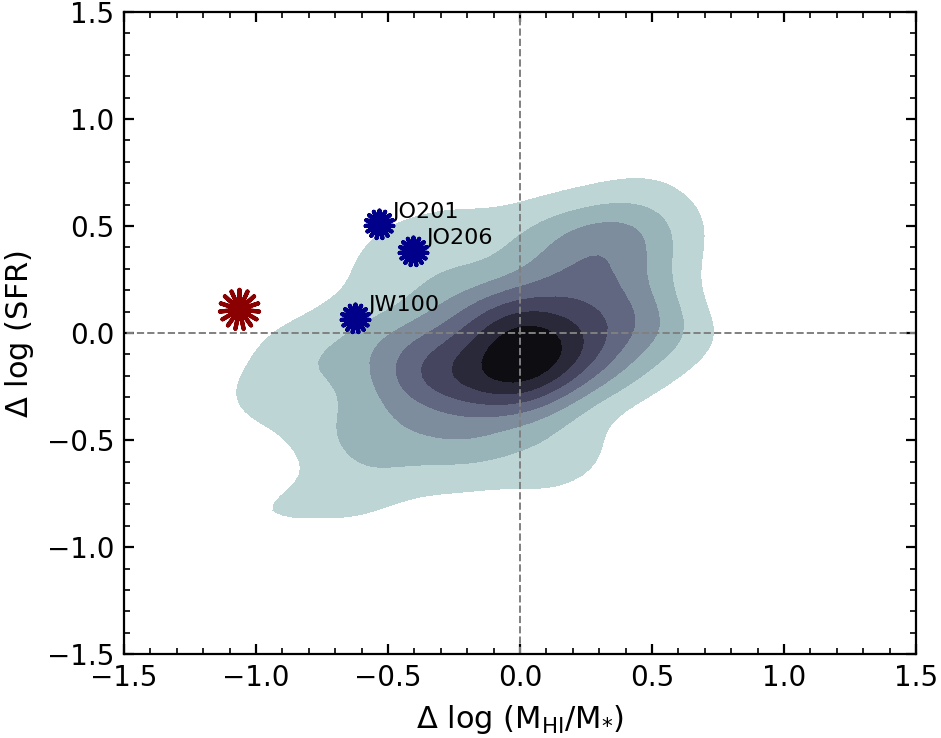}\hfill
  \includegraphics[width=\columnwidth]{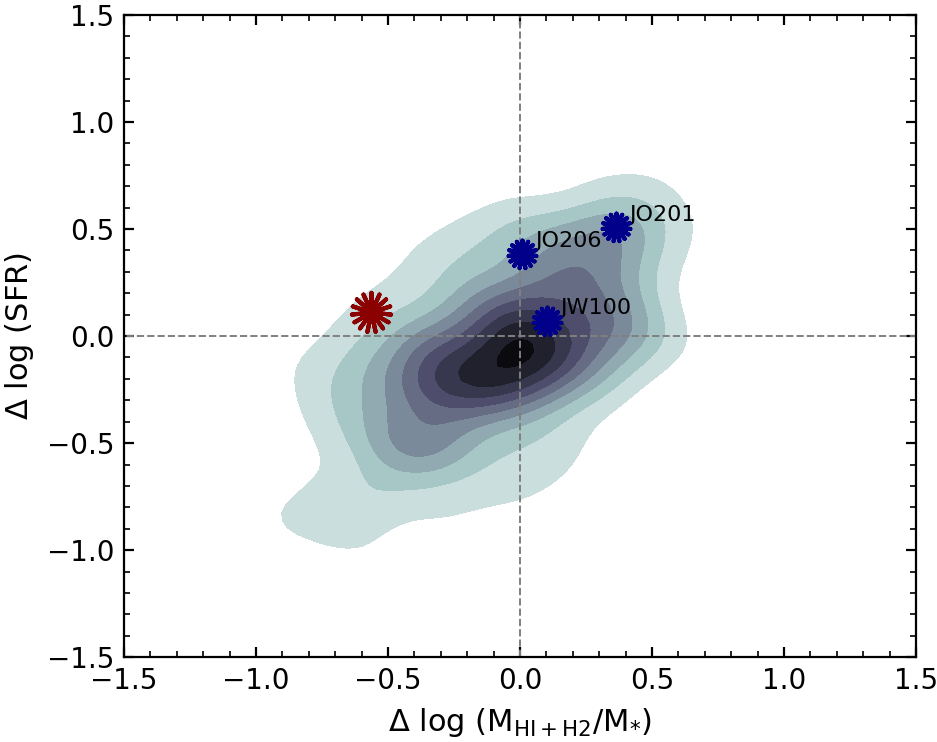}\hfill
    \caption{Star formation rate offset from the main sequence as a function of the \HI\ gas (top) and total gas (bottom) fraction offset from the xGASS gas scaling relations. Grey contours are `normal' field galaxies. The red asterisk represents ESO\,137-001, and the blue asterisks show the comparison sample of jellyfish galaxies.}\label{offset}
\end{figure}

In Figure \ref{offset}, the deviation of the observed star formation rate (SFR) from the main sequence is shown, defined as $\Delta$SFR = log(SFR/SFR$_{\rm ms}$), and plotted as a function of the deviation of the measured gas fraction ($f_{\rm gas}$ = M$_{\rm gas}$/M$_{\ast}$) from the expected, $\Delta f_{\rm gas}$ = log($f_{\rm gas}$/$f_{\rm gas, exp}$). We adopt the SFR$_{\rm ms}$ described by \citet{Saintonge2016};
\begin{equation}
\mathrm{logSFR = -2.332x + 0.415x^{2} - 0.01828x^{3}}   
\end{equation}

where $x$ = log(M$_{\ast}$/M$_{\odot}$) and main sequence galaxies are those with |$\Delta$SFR| $< 0.4$. The expected \HI\ and total gas fraction were derived from the xGASS scaling relations (i.e. Figs.\,6 \& 8 in \citealp{Catinella2018}). In this $\Delta$SFR-$\Delta f_{\rm gas}$ parameter space, the field (`normal') galaxies are on the main sequence, concentrated at ($\Delta$SFR, $\Delta f_{\rm gas}$) $\approx$ (0,0) and are distributed such that the SFR-offset positively correlates with the observed gas fraction offset.
The top panel of Fig.\,\ref{offset} shows $\Delta$SFR as a function of the \HI\ gas fraction offset $\Delta f_{\rm HI}$. In this panel, ESO\,137-001 and the comparison jellyfish galaxies lie in the top left quadrant. Although their SFR offset is within the scatter of the main sequence, ESO\,137-001 and the other jellyfish galaxies are at the upper edge of the distribution, which indicates a significant excess SFR for their measured \HI\ mass. Indeed, we measure a \HI\ depletion time, $\tau_{dep}$ = M$_{\rm HI}$/SFR, of 0.29 Gyr for this galaxy which is much lower than the 2 Gyr value typical for star forming galaxies with a similar stellar mass.
In the bottom panel, we show $\Delta$SFR as a function of the total (\HI\ + H$_{2}$) gas fraction deviation, $\Delta f_{\rm totgas}$. In this plot ESO\,137-001 remains in the top left quadrant with a reduced but still significant offset of log$\Delta f_{\rm totgas}$ = -0.56. This is in contrast with the comparison jellyfish galaxies, which have moved to the top right quadrant and are within the scatter of main-sequence galaxies, \citet{Moretti2020, Moretti2023} suggested that this indicates enhanced \HI\ to H$_2$ conversion during peak stripping. Instead, the position of ESO\,137-001 on this plot suggests that this galaxy may be characterised more by efficient \HI\ stripping than by enhanced \HI\ to H$_2$ conversion. Given its measured star formation rate (SFR) of $1.2 \, M_{\odot}/\text{yr}$, to lie precisely on the main sequence, the galaxy would be expected to have a molecular hydrogen mass $M_{H_2} = 2.4 \times 10^9 \, M_{\odot}$ consistent with measurements from ALMA observations \citep{Jachym2019}.  

\begin{figure}
   \centering
 \includegraphics[width=\columnwidth]{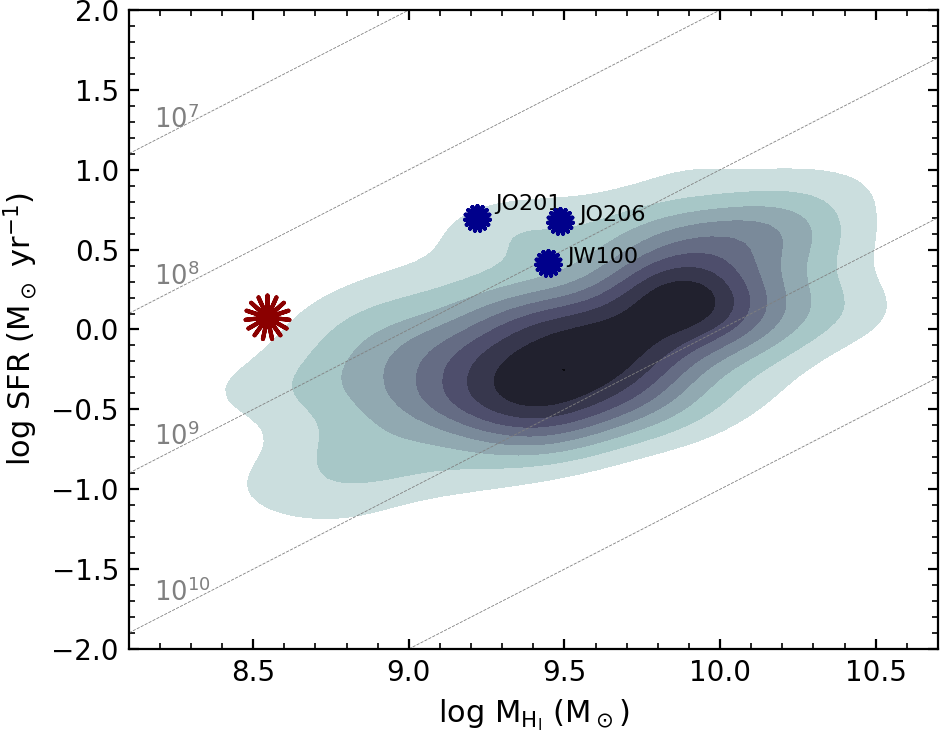}\hfill
    \caption{Star formation rate as function of H\textsc{i} mass for galaxies with similar stellar masses as ESO\,137-001 and jellyfish counterparts. Grey contours are galaxies from xGASS. The red asterisk represents ESO\,137-001 while the blue points are jellyfish galaxies, JO206, JO201 and JW100.}\label{figsfr}
 \end{figure}

ESO\,137-001 appears to occupy a slightly different region of parameter space in Fig.\,\ref{offset}. It has lost much more of its \HI\ compared to its counterparts (see \citealp{Ramatsoku2020, Deb2022}). It also has a much lower SFR than its jellyfish counterparts, as shown in Fig.\,\ref{figsfr}. There, we plot the control field sample selected xGASS, consisting of galaxies with stellar masses comparable to ESO\,137-001 (M$_{\ast} \sim 1.4 \times\ 10^{10}$ \MSUN; \citealp{Sivanandam2010}) spanning one order of magnitude. We also show the jellyfish counterparts. The significant \HI\ deficiency and lower SFR of ESO 137-001 compared to JO201 and JO206, which have a similar stellar mass, indicate that it is at a somewhat more advanced stage in its evolution or that it has experienced increased levels of ram-pressure stripping.  %

\begin{figure}[h]
   \centering
 \includegraphics[width=\columnwidth]{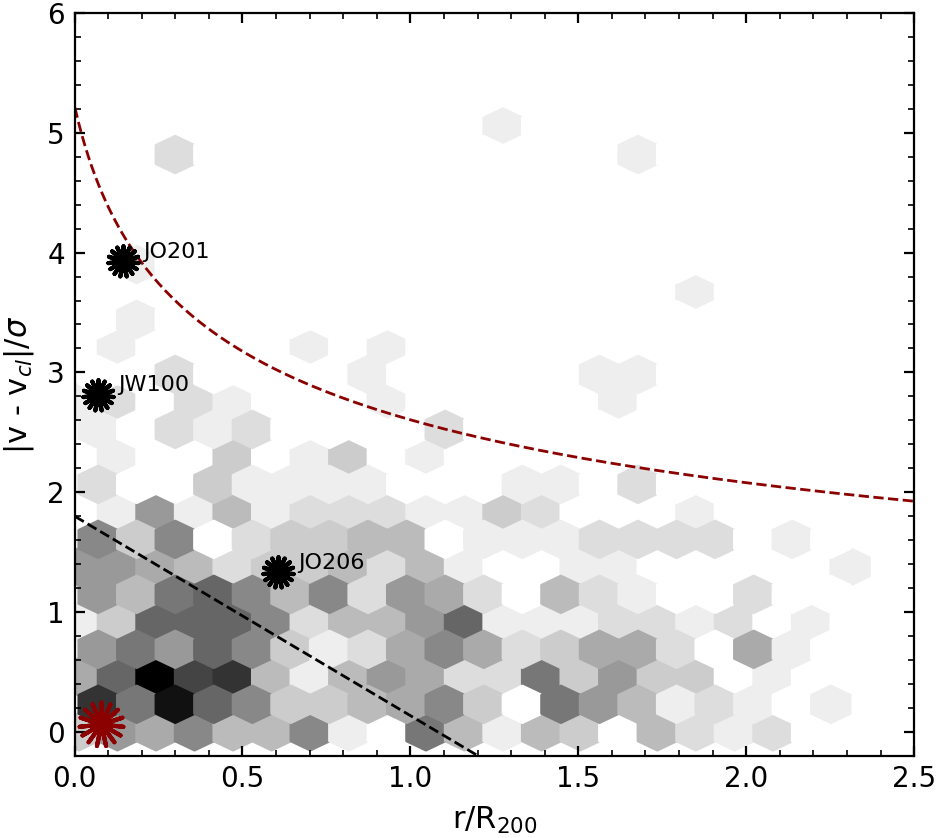}
    \caption{Composite projected phase-space diagram of spectroscopically confirmed galaxy members of clusters Abell 85, IIZW108, Abell 2626 and ACO\,3627 within which the comparison jellyfish galaxies (JO201, JO206, JW100) and ESO\,137-001 are located, respectively. The red curve is the escape velocity in an NFW halo \citep{Navarro1996}. The dashed black line indicates the region where galaxies are most likely virialised. Black asterisks are the comparison sample of jellyfish galaxies within their respective galaxy clusters, and the red asterisk represents ESO\,137-001.}\label{phasespace}
 \end{figure}

In Fig.\,\ref{phasespace}, we compare the inferred orbital histories or infall times of ESO\,137-001 and other jellyfish galaxies within their respective clusters on the phase-space diagram \citep{Jaffe2015, Rhee2017}. This plot shows the projected distance from the cluster centre versus the galaxy velocities relative to the cluster. Galaxies fall from large radii at high velocity and eventually settle into the cluster central (i.e. virialised) region delineated by the grey dashed line (\citealp{Mahajan2011, Jaffe2015}). The expected ram pressure increases with the ICM density, which increases closer to the cluster centre and with the square of the differential velocity. Jellyfish galaxies are found outside the virialised region at $r/R\_{200} \lesssim 0.6$ travelling at high velocities $\delta v \geq 1.5\sigma$. In contrast, ESO\,137-001 appears to be located within the virialised region. However, this apparent position largely reflects its motion in the plane of the sky. Like its counterparts, ESO\,137-001 is likely falling into the cluster for the first time and is approaching the pericentre.

\section{Summary}\label{summary}
We conducted \HI\ observations of the archetypical jellyfish galaxy, ESO\,137-001, located in the nearby massive galaxy cluster, ACO\,3627 at $z$ = 0.01625. It lies near the centre of ACO\,3627 at a projected distance of 200 kpc from its centre. ESO\,137-001 is undergoing ram-pressure stripping as it falls into the cluster at velocities of $\sim3000$ \kms. Prior ATCA observations did not detect the presence of \HI\ in the galaxy to the upper limit of M$_{\rm HI}$ $\sim 6.0 \times 10^{8}$\,\MSUN\ at the 3$\sigma$ level for a 150\kms\ linewidth. The MeerKAT observations enabled the detection of \HI\ in the main body and extended tail of the galaxy for the first time.

We detected \HI\ in the disc and outer regions and measured a total \HI\ mass, M$_{\rm HI} = (3.5 \pm 0.4) \times 10^{8}$ \MSUN. Our analysis revealed that ESO\,137-001 is severely \HI\ deficient, having lost $\sim$90\% of its initial \HI\ mass; about 2/3 (M$_{\rm HI} = 2.3 \times 10^{8}$ \MSUN) of the surviving \HI\ is found at a larger radius than expected for a normal gas disc and forms a 40\,kpc tail coincident with the tail detected at other wavelengths (e.g. X-ray, CO, H$\alpha$).
We also find that $\sim10$\% of the surviving \HI\ is still found within the stellar disc, consistent with the expectation in case of an outside-in truncation due to ram pressure.

ESO\,137-001 has a high star formation rate for the amount of \HI\ detected. Our analysis of the SFR in relation to the total gas fraction indicates that ESO\,137-001 lies on the upper edge but within the scatter of the SFR main sequence, while the jellyfish galaxies under comparison are situated on the main sequence; the latter typically indicate enhanced \HI-to-H$_{\rm 2}$ conversion in jellyfish galaxies during peak ram-pressure stripping reported in recent studies. However, ESO\,137-001 appears to be more characterised by efficient \HI\ stripping than enhanced \HI-to-H$_{\rm 2}$ conversion, which suggests it might be in an advanced evolutionary stripping phase.

\begin{acknowledgements}
MR's research is supported by the SARAO HCD programme via the `New Scientific Frontiers with Precision Radio Interferometry II' research group grant. OMS's research is supported by the South African Research Chairs Initiative of the Department of Science and Technology and the National Research Foundation (grant No. 81737). MS acknowledges the support from the NSF grant 2407821. The authors thank Pavel J'achym for providing the ALMA CO image. The MeerKAT telescope is operated by the South African Radio Astronomy Observatory, which is a facility of the National Research Foundation, an agency of the Department of Science and Innovation.\\
\indent Part of the data published here have been reduced using the CARACal pipeline, partially supported by ERC Starting grant number 679629 “FORNAX”, MAECI Grant Number ZA18GR02, DST-NRF Grant Number 113121 as part of the ISARP Joint Research Scheme, and BMBF project 05A17PC2 for D-MeerKAT. Information about CARACal can be obtained online under the URL:: \url{https://caracal.readthedocs.io}.\\
\noindent We acknowledge the use of computing facilities of the Inter-University Institute for Data Intensive Astronomy (IDIA) for part of this work. IDIA is a partnership of the Universities of Cape Town, of the Western Cape and of Pretoria.
\end{acknowledgements}

\bibliographystyle{aa} 
\bibliography{ESORefv2.bib} 

\begin{thebibliography}{102}
\expandafter\ifx\csname natexlab\endcsname\relax\def\natexlab#1{#1}\fi

\bibitem[{{Abell} {et~al.}(1989){Abell}, {Corwin}, \& {Olowin}}]{Abell1989}
{Abell}, G.~O., {Corwin}, Jr., H.~G., \& {Olowin}, R.~P. 1989, \apjs, 70, 1

\bibitem[{{Asai} {et~al.}(2005){Asai}, {Fukuda}, \& {Matsumoto}}]{Asai2005}
{Asai}, N., {Fukuda}, N., \& {Matsumoto}, R. 2005, Advances in Space Research,
  36, 636

\bibitem[{{Boselli} {et~al.}(2018){Boselli}, {Fossati}, {Ferrarese},
  {Boissier}, {Consolandi}, {Longobardi}, {Amram}, {Balogh}, {Barmby},
  {Boquien}, {Boulanger}, {Braine}, {Buat}, {Burgarella}, {Combes}, {Contini},
  {Cortese}, {C{\^o}t{\'e}}, {C{\^o}t{\'e}}, {Cuillandre}, {Drissen}, {Epinat},
  {Fumagalli}, {Gallagher}, {Gavazzi}, {Gomez-Lopez}, {Gwyn}, {Harris},
  {Hensler}, {Koribalski}, {Marcelin}, {McConnachie}, {Miville-Deschenes},
  {Navarro}, {Patton}, {Peng}, {Plana}, {Prantzos}, {Robert}, {Roediger},
  {Roehlly}, {Russeil}, {Salome}, {Sanchez-Janssen}, {Serra}, {Spekkens},
  {Sun}, {Taylor}, {Tonnesen}, {Vollmer}, {Willis}, {Wozniak}, {Burdullis},
  {Devost}, {Mahoney}, {Manset}, {Petric}, {Prunet}, \&
  {Withington}}]{Boselli2018}
{Boselli}, A., {Fossati}, M., {Ferrarese}, L., {et~al.} 2018, \aap, 614, A56

\bibitem[{{Brown} {et~al.}(2015){Brown}, {Catinella}, {Cortese}, {Kilborn},
  {Haynes}, \& {Giovanelli}}]{Brown2015}
{Brown}, T., {Catinella}, B., {Cortese}, L., {et~al.} 2015, \mnras, 452, 2479

\bibitem[{{Brown} {et~al.}(2023){Brown}, {Roberts}, {Thorp}, {Ellison},
  {Zabel}, {Wilson}, {Bah{\'e}}, {Bisaria}, {Bolatto}, {Boselli}, {Chung},
  {Cortese}, {Catinella}, {Davis}, {Jim{\'e}nez-Donaire}, {Lagos}, {Lee},
  {Parker}, {Smith}, {Spekkens}, {Stevens}, {Villanueva}, \&
  {Watts}}]{Brown2023}
{Brown}, T., {Roberts}, I.~D., {Thorp}, M., {et~al.} 2023, \apj, 956, 37

\bibitem[{{Catinella} {et~al.}(2018){Catinella}, {Saintonge}, {Janowiecki},
  {Cortese}, {Dav{\'e}}, {Lemonias}, {Cooper}, {Schiminovich}, {Hummels},
  {Fabello}, {Ger{\'e}b}, {Kilborn}, \& {Wang}}]{Catinella2018}
{Catinella}, B., {Saintonge}, A., {Janowiecki}, S., {et~al.} 2018, \mnras, 476,
  875

\bibitem[{{Chabrier}(2003)}]{Chabrier2003}
{Chabrier}, G. 2003, \pasp, 115, 763

\bibitem[{{Chen} {et~al.}(2020){Chen}, {Sun}, {Yagi}, {Bravo-Alfaro}, {Brinks},
  {Kenney}, {Combes}, {Sivanandam}, {Jachym}, {Fossati}, {Gavazzi}, {Boselli},
  {Nulsen}, {Sarazin}, {Ge}, {Yoshida}, \& {Roediger}}]{Chen2020}
{Chen}, H., {Sun}, M., {Yagi}, M., {et~al.} 2020, \mnras, 496, 4654

\bibitem[{{Chung} {et~al.}(2009){Chung}, {van Gorkom}, {Kenney}, {Crowl}, \&
  {Vollmer}}]{Chung2009}
{Chung}, A., {van Gorkom}, J.~H., {Kenney}, J.~D.~P., {Crowl}, H., \&
  {Vollmer}, B. 2009, \aj, 138, 1741

\bibitem[{{Chung} {et~al.}(2007){Chung}, {van Gorkom}, {Kenney}, \&
  {Vollmer}}]{Chung2007}
{Chung}, A., {van Gorkom}, J.~H., {Kenney}, J. D.~P., \& {Vollmer}, B. 2007,
  \apjl, 659, L115

\bibitem[{{Cramer} {et~al.}(2020){Cramer}, {Kenney}, {Cortes}, {Cortes P.~C.},
  {Vlahakis}, {J{\'a}chym}, {Pompei}, \& {Rubio}}]{Cramer2020}
{Cramer}, W.~J., {Kenney}, J.~D.~P., {Cortes}, J.~R., {et~al.} 2020, \apj, 901,
  95

\bibitem[{{Cramer} {et~al.}(2019){Cramer}, {Kenney}, {Sun}, {Crowl}, {Yagi},
  {J{\'a}chym}, {Roediger}, \& {Waldron}}]{Cramer2019}
{Cramer}, W.~J., {Kenney}, J.~D.~P., {Sun}, M., {et~al.} 2019, \apj, 870, 63

\bibitem[{{Crowl} {et~al.}(2005){Crowl}, {Kenney}, {van Gorkom}, \&
  {Vollmer}}]{Crowl2005}
{Crowl}, H.~H., {Kenney}, J. D.~P., {van Gorkom}, J.~H., \& {Vollmer}, B. 2005,
  \aj, 130, 65

\bibitem[{{Deb} {et~al.}(2020){Deb}, {Verheijen}, {Gullieuszik}, {Poggianti},
  {van Gorkom}, {Ramatsoku}, {Serra}, {Moretti}, {Vulcani}, {Bettoni},
  {Jaff{\'e}}, {Tonnesen}, \& {Fritz}}]{Deb2020}
{Deb}, T., {Verheijen}, M. A.~W., {Gullieuszik}, M., {et~al.} 2020, \mnras,
  494, 5029

\bibitem[{{Deb} {et~al.}(2022){Deb}, {Verheijen}, {Poggianti}, {Moretti}, {van
  der Hulst}, {Vulcani}, {Ramatsoku}, {Serra}, {Healy}, {Gullieuszik},
  {Bacchini}, {Ignesti}, {M{\"u}ller}, {Zabel}, {Luber}, {Jaff{\"e}}, \&
  {Gitti}}]{Deb2022}
{Deb}, T., {Verheijen}, M. A.~W., {Poggianti}, B.~M., {et~al.} 2022, \mnras,
  516, 2683

\bibitem[{{D{\'e}nes} {et~al.}(2014){D{\'e}nes}, {Kilborn}, \&
  {Koribalski}}]{Denes2014}
{D{\'e}nes}, H., {Kilborn}, V.~A., \& {Koribalski}, B.~S. 2014, \mnras, 444,
  667

\bibitem[{{Dickey}(1997)}]{Dickey1997}
{Dickey}, J.~M. 1997, \aj, 113, 1939

\bibitem[{{Dressler}(1980)}]{Dressler1980}
{Dressler}, A. 1980, \apj, 236, 351

\bibitem[{{Dressler} {et~al.}(1987){Dressler}, {Faber}, {Burstein}, {Davies},
  {Lynden-Bell}, {Terlevich}, \& {Wegner}}]{Dressler1987GA}
{Dressler}, A., {Faber}, S.~M., {Burstein}, D., {et~al.} 1987, \apjl, 313, L37

\bibitem[{{Ebeling} {et~al.}(2014){Ebeling}, {Stephenson}, \&
  {Edge}}]{Ebeling2014}
{Ebeling}, H., {Stephenson}, L.~N., \& {Edge}, A.~C. 2014, \apjl, 781, L40

\bibitem[{{Eckert} {et~al.}(2017){Eckert}, {Gaspari}, {Owers}, {Roediger},
  {Molendi}, {Gastaldello}, {Paltani}, {Ettori}, {Venturi}, {Rossetti}, \&
  {Rudnick}}]{Eckert2017}
{Eckert}, D., {Gaspari}, M., {Owers}, M.~S., {et~al.} 2017, \aap, 605, A25

\bibitem[{{Ettori} \& {Fabian}(2000)}]{Ettori2000}
{Ettori}, S. \& {Fabian}, A.~C. 2000, \mnras, 317, L57

\bibitem[{{Foltz} {et~al.}(2018){Foltz}, {Wilson}, {Muzzin}, {Cooper},
  {Nantais}, {van der Burg}, {Cerulo}, {Chan}, {Fillingham}, {Surace}, {Webb},
  {Noble}, {Lacy}, {McDonald}, {Rudnick}, {Lidman}, {Demarco},
  {Hlavacek-Larrondo}, {Yee}, {Perlmutter}, \& {Hayden}}]{Foltz2018}
{Foltz}, R., {Wilson}, G., {Muzzin}, A., {et~al.} 2018, \apj, 866, 136

\bibitem[{{Fossati} {et~al.}(2016){Fossati}, {Fumagalli}, {Boselli}, {Gavazzi},
  {Sun}, \& {Wilman}}]{Fossati2016}
{Fossati}, M., {Fumagalli}, M., {Boselli}, A., {et~al.} 2016, \mnras, 455, 2028

\bibitem[{{Fossati} {et~al.}(2012){Fossati}, {Gavazzi}, {Boselli}, \&
  {Fumagalli}}]{Fossati2012}
{Fossati}, M., {Gavazzi}, G., {Boselli}, A., \& {Fumagalli}, M. 2012, \aap,
  544, A128

\bibitem[{{Fossati} {et~al.}(2018){Fossati}, {Mendel}, {Boselli}, {Cuillandre},
  {Vollmer}, {Boissier}, {Consolandi}, {Ferrarese}, {Gwyn}, {Amram}, {Boquien},
  {Buat}, {Burgarella}, {Cortese}, {C{\^o}t{\'e}}, {C{\^o}t{\'e}}, {Durrell},
  {Fumagalli}, {Gavazzi}, {Gomez-Lopez}, {Hensler}, {Koribalski}, {Longobardi},
  {Peng}, {Roediger}, {Sun}, \& {Toloba}}]{Fossati2018}
{Fossati}, M., {Mendel}, J.~T., {Boselli}, A., {et~al.} 2018, \aap, 614, A57

\bibitem[{{Frater} {et~al.}(1992){Frater}, {Brooks}, \&
  {Whiteoak}}]{Frater1992}
{Frater}, R.~H., {Brooks}, J.~W., \& {Whiteoak}, J.~B. 1992, Journal of
  Electrical and Electronics Engineering Australia, 12, 103

\bibitem[{{Fumagalli} {et~al.}(2014){Fumagalli}, {Fossati}, {Hau}, {Gavazzi},
  {Bower}, {Sun}, \& {Boselli}}]{Fumagalli2014}
{Fumagalli}, M., {Fossati}, M., {Hau}, G.~K.~T., {et~al.} 2014, \mnras, 445,
  4335

\bibitem[{{Gavazzi}(1989)}]{Gavazzi1989}
{Gavazzi}, G. 1989, \apj, 346, 59

\bibitem[{{Gavazzi} {et~al.}(1995){Gavazzi}, {Contursi}, {Carrasco}, {Boselli},
  {Kennicutt}, {Scodeggio}, \& {Jaffe}}]{Gavazzi1995}
{Gavazzi}, G., {Contursi}, A., {Carrasco}, L., {et~al.} 1995, \aap, 304, 325

\bibitem[{{G{\"o}ller} {et~al.}(2023){G{\"o}ller}, {Joshi}, {Rohr}, {Zinger},
  \& {Pillepich}}]{Goller2023}
{G{\"o}ller}, J., {Joshi}, G.~D., {Rohr}, E., {Zinger}, E., \& {Pillepich}, A.
  2023, \mnras, 525, 3551

\bibitem[{{Grobler} {et~al.}(2014){Grobler}, {Nunhokee}, {Smirnov}, {van Zyl},
  \& {de Bruyn}}]{Grobler2014}
{Grobler}, T.~L., {Nunhokee}, C.~D., {Smirnov}, O.~M., {van Zyl}, A.~J., \& {de
  Bruyn}, A.~G. 2014, \mnras, 439, 4030

\bibitem[{{Gunn} \& {Gott}(1972)}]{Gunn1972}
{Gunn}, J.~E. \& {Gott}, III, J.~R. 1972, \apj, 176, 1

\bibitem[{{Haynes} {et~al.}(1984){Haynes}, {Giovanelli}, \&
  {Chincarini}}]{Haynes1984}
{Haynes}, M.~P., {Giovanelli}, R., \& {Chincarini}, G.~L. 1984, \araa, 22, 445

\bibitem[{{Jachym} {et~al.}(2014){Jachym}, {Combes}, {Cortese}, {Sun}, \&
  {Kenney}}]{Jachym2014}
{Jachym}, P., {Combes}, F., {Cortese}, L., {Sun}, M., \& {Kenney}, J.~D.~P.
  2014, \apj, 792, 11

\bibitem[{{J{\'a}chym} {et~al.}(2019){J{\'a}chym}, {Kenney}, {Sun}, {Combes},
  {Cortese}, {Scott}, {Sivanandam}, {Brinks}, {Roediger}, {Palou{\v{s}}}, \&
  {Fumagalli}}]{Jachym2019}
{J{\'a}chym}, P., {Kenney}, J. D.~P., {Sun}, M., {et~al.} 2019, \apj, 883, 145

\bibitem[{{J{\'a}chym} {et~al.}(2017){J{\'a}chym}, {Sun}, {Kenney}, {Cortese},
  {Combes}, {Yagi}, {Yoshida}, {Palou{\v{s}}}, \& {Roediger}}]{Jachym2017}
{J{\'a}chym}, P., {Sun}, M., {Kenney}, J. D.~P., {et~al.} 2017, \apj, 839, 114

\bibitem[{{Jaff{\'e}} {et~al.}(2015){Jaff{\'e}}, {Smith}, {Candlish},
  {Poggianti}, {Sheen}, \& {Verheijen}}]{Jaffe2015}
{Jaff{\'e}}, Y.~L., {Smith}, R., {Candlish}, G.~N., {et~al.} 2015, \mnras, 448,
  1715

\bibitem[{{Jarrett} {et~al.}(2000){Jarrett}, {Chester}, {Cutri}, {Schneider},
  {Skrutskie}, \& {Huchra}}]{Jarrett2000}
{Jarrett}, T.~H., {Chester}, T., {Cutri}, R., {et~al.} 2000, \aj, 119, 2498

\bibitem[{{Jonas} \& {MeerKAT Team}(2016)}]{Jonas2016}
{Jonas}, J. \& {MeerKAT Team}. 2016, in MeerKAT Science: On the Pathway to the
  SKA, 1

\bibitem[{{J{\'o}zsa} {et~al.}(2022){J{\'o}zsa}, {Andati}, {de Blok}, {Hugo},
  {Kleiner}, {Kamphuis}, {Moln{\'a}r}, {Makhathini}, {Maccagni}, {Perkins},
  {Ramaila}, {Ramatsoku}, {Serra}, {Smirnov}, {Thorat}, \& {White}}]{Josza2022}
{J{\'o}zsa}, G. I.~G., {Andati}, L. A.~L., {de Blok}, W.~J.~G., {et~al.} 2022,
  in Astronomical Society of the Pacific Conference Series, Vol. 532,
  Astronomical Society of the Pacific Conference Series, ed. J.~E. {Ruiz},
  F.~{Pierfedereci}, \& P.~{Teuben}, 447

\bibitem[{{Kapferer} {et~al.}(2009){Kapferer}, {Sluka}, {Schindler}, {Ferrari},
  \& {Ziegler}}]{Kapferer2009}
{Kapferer}, W., {Sluka}, C., {Schindler}, S., {Ferrari}, C., \& {Ziegler}, B.
  2009, \aap, 499, 87

\bibitem[{{Kenney} {et~al.}(2004){Kenney}, {van Gorkom}, \&
  {Vollmer}}]{Kenney2004}
{Kenney}, J.~D.~P., {van Gorkom}, J.~H., \& {Vollmer}, B. 2004, \aj, 127, 3361

\bibitem[{{Kenyon} {et~al.}(2018){Kenyon}, {Smirnov}, {Grobler}, \&
  {Perkins}}]{Kenyon2018}
{Kenyon}, J.~S., {Smirnov}, O.~M., {Grobler}, T.~L., \& {Perkins}, S.~J. 2018,
  \mnras, 478, 2399

\bibitem[{{Koribalski} {et~al.}(2024){Koribalski}, {Duchesne}, {Lenc},
  {Venturi}, {Botteon}, {Shabala}, {Vernstrom}, {Carretti}, {Norris},
  {Anderson}, {Hopkins}, {Riseley}, {Gupta}, \& {Velovi{\'c}}}]{Koribalski2024}
{Koribalski}, B.~S., {Duchesne}, S.~W., {Lenc}, E., {et~al.} 2024, \mnras, 533,
  608

\bibitem[{{Kronberger} {et~al.}(2008){Kronberger}, {Kapferer}, {Ferrari},
  {Unterguggenberger}, \& {Schindler}}]{Kronberger2008}
{Kronberger}, T., {Kapferer}, W., {Ferrari}, C., {Unterguggenberger}, S., \&
  {Schindler}, S. 2008, \aap, 481, 337

\bibitem[{{Lee} {et~al.}(2017){Lee}, {Chung}, {Tonnesen}, {Kenney}, {Wong},
  {Vollmer}, {Petitpas}, {Crowl}, \& {van Gorkom}}]{Lee2017}
{Lee}, B., {Chung}, A., {Tonnesen}, S., {et~al.} 2017, \mnras, 466, 1382

\bibitem[{{Loni} {et~al.}(2021){Loni}, {Serra}, {Kleiner}, {Cortese},
  {Catinella}, {Koribalski}, {Jarrett}, {Molnar}, {Davis}, {Iodice},
  {Lee-Waddell}, {Loi}, {Maccagni}, {Peletier}, {Popping}, {Ramatsoku},
  {Smith}, \& {Zabel}}]{Loni2021}
{Loni}, A., {Serra}, P., {Kleiner}, D., {et~al.} 2021, \aap, 648, A31

\bibitem[{{Luo} {et~al.}(2023){Luo}, {Sun}, {J{\'a}chym}, {Waldron}, {Fossati},
  {Fumagalli}, {Boselli}, {Combes}, {Kenney}, {Li}, \& {Gronke}}]{Luo2023}
{Luo}, R., {Sun}, M., {J{\'a}chym}, P., {et~al.} 2023, \mnras, 521, 6266

\bibitem[{{Mahajan} {et~al.}(2011){Mahajan}, {Mamon}, \&
  {Raychaudhury}}]{Mahajan2011}
{Mahajan}, S., {Mamon}, G.~A., \& {Raychaudhury}, S. 2011, \mnras, 416, 2882

\bibitem[{Makhathini(2018)}]{makhathini2018}
Makhathini, S. 2018, PhD thesis, Rhodes University, Drosty Rd, Grahamstown,
  6139, Eastern Cape, South Africa, available via
  \url{http://hdl.handle.net/10962/57348}

\bibitem[{{Martinsson} {et~al.}(2016){Martinsson}, {Verheijen}, {Bershady},
  {Westfall}, {Andersen}, \& {Swaters}}]{Martinsson2016}
{Martinsson}, T.~P.~K., {Verheijen}, M.~A.~W., {Bershady}, M.~A., {et~al.}
  2016, \aap, 585, A99

\bibitem[{{Mauch} {et~al.}(2020){Mauch}, {Cotton}, {Condon}, {Matthews},
  {Abbott}, {Adam}, {Aldera}, {Asad}, {Bauermeister}, {Bennett}, {Bester},
  {Botha}, {Brederode}, {Brits}, {Buchner}, {Burger}, {Camilo}, {Chalmers},
  {Cheetham}, {de Villiers}, {de Villiers}, {Dikgale-Mahlakoana}, {du Toit},
  {Esterhuyse}, {Fadana}, {Fanaroff}, {Fataar}, {February}, {Frank},
  {Gamatham}, {Geyer}, {Goedhart}, {Gounden}, {Gumede}, {Heywood}, {Hlakola},
  {Horrell}, {Hugo}, {Isaacson}, {J{\'o}zsa}, {Jonas}, {Julie}, {Kapp},
  {Kasper}, {Kenyon}, {Kotz{\'e}}, {Kriek}, {Kriel}, {Kusel}, {Lehmensiek},
  {Loots}, {Lord}, {Lunsky}, {Madisa}, {Magnus}, {Main}, {Malan}, {Manley},
  {Marais}, {Martens}, {Merry}, {Millenaar}, {Mnyandu}, {Moeng}, {Mokone},
  {Monama}, {Mphego}, {New}, {Ngcebetsha}, {Ngoasheng}, {Ockards}, {Oozeer},
  {Otto}, {Patel}, {Peens-Hough}, {Perkins}, {Ramaila}, {Ramudzuli}, {Renil},
  {Richter}, {Robyntjies}, {Salie}, {Schollar}, {Schwardt}, {Serylak},
  {Siebrits}, {Sirothia}, {Smirnov}, {Sofeya}, {Stone}, {Taljaard}, {Tasse},
  {Theron}, {Tiplady}, {Toruvanda}, {Twum}, {van Balla}, {van der Byl}, {van
  der Merwe}, {Van Tonder}, {Wallace}, {Welz}, {Williams}, \&
  {Xaia}}]{Mauch2020}
{Mauch}, T., {Cotton}, W.~D., {Condon}, J.~J., {et~al.} 2020, \apj, 888, 61

\bibitem[{{McMullin} {et~al.}(2007){McMullin}, {Waters}, {Schiebel}, {Young},
  \& {Golap}}]{McMullin2007}
{McMullin}, J.~P., {Waters}, B., {Schiebel}, D., {Young}, W., \& {Golap}, K.
  2007, in Astronomical Society of the Pacific Conference Series, Vol. 376,
  Astronomical Data Analysis Software and Systems XVI, ed. R.~A. {Shaw},
  F.~{Hill}, \& D.~J. {Bell}, 127

\bibitem[{{Meyer} {et~al.}(2017){Meyer}, {Robotham}, {Obreschkow}, {Westmeier},
  {Duffy}, \& {Staveley-Smith}}]{Meyer2017}
{Meyer}, M., {Robotham}, A., {Obreschkow}, D., {et~al.} 2017, \pasa, 34, 52

\bibitem[{{Moretti} {et~al.}(2020){Moretti}, {Paladino}, {Poggianti}, {Serra},
  {Ramatsoku}, {Franchetto}, {Deb}, {Gullieuszik}, {Tomi{\v{c}}i{\'c}},
  {Mingozzi}, {Vulcani}, {Radovich}, {Bettoni}, \& {Fritz}}]{Moretti2020}
{Moretti}, A., {Paladino}, R., {Poggianti}, B.~M., {et~al.} 2020, \apjl, 897,
  L30

\bibitem[{{Moretti} {et~al.}(2023){Moretti}, {Serra}, {Bacchini}, {Paladino},
  {Ramatsoku}, {Poggianti}, {Vulcani}, {Deb}, {Gullieuszik}, {Fritz}, \&
  {Wolter}}]{Moretti2023}
{Moretti}, A., {Serra}, P., {Bacchini}, C., {et~al.} 2023, \apj, 955, 153

\bibitem[{{M{\"u}ller} {et~al.}(2021){M{\"u}ller}, {Poggianti}, {Pfrommer},
  {Adebahr}, {Serra}, {Ignesti}, {Sparre}, {Gitti}, {Dettmar}, {Vulcani}, \&
  {Moretti}}]{Muller2021}
{M{\"u}ller}, A., {Poggianti}, B.~M., {Pfrommer}, C., {et~al.} 2021, Nature
  Astronomy, 5, 159

\bibitem[{{Mun} {et~al.}(2021){Mun}, {Hwang}, {Lee}, {Chung}, {Yoon}, \&
  {Lee}}]{Mun2021}
{Mun}, J.~Y., {Hwang}, H.~S., {Lee}, M.~G., {et~al.} 2021, Journal of Korean
  Astronomical Society, 54, 17

\bibitem[{{Navarro} {et~al.}(1996){Navarro}, {Frenk}, \& {White}}]{Navarro1996}
{Navarro}, J.~F., {Frenk}, C.~S., \& {White}, S. D.~M. 1996, \apj, 462, 563

\bibitem[{{Nelson} {et~al.}(2019){Nelson}, {Pillepich}, {Springel}, {Pakmor},
  {Weinberger}, {Genel}, {Torrey}, {Vogelsberger}, {Marinacci}, \&
  {Hernquist}}]{Nelson2019}
{Nelson}, D., {Pillepich}, A., {Springel}, V., {et~al.} 2019, \mnras, 490, 3234

\bibitem[{{Offringa} {et~al.}(2010){Offringa}, {de Bruyn}, {Biehl}, {Zaroubi},
  {Bernardi}, \& {Pandey}}]{Offringa2010}
{Offringa}, A.~R., {de Bruyn}, A.~G., {Biehl}, M., {et~al.} 2010, \mnras, 405,
  155

\bibitem[{{Offringa} {et~al.}(2014){Offringa}, {McKinley}, {Hurley-Walker},
  {Briggs}, {Wayth}, {Kaplan}, {Bell}, {Feng}, {Neben}, {Hughes}, {Rhee},
  {Murphy}, {Bhat}, {Bernardi}, {Bowman}, {Cappallo}, {Corey}, {Deshpande},
  {Emrich}, {Ewall-Wice}, {Gaensler}, {Goeke}, {Greenhill}, {Hazelton},
  {Hindson}, {Johnston-Hollitt}, {Jacobs}, {Kasper}, {Kratzenberg}, {Lenc},
  {Lonsdale}, {Lynch}, {McWhirter}, {Mitchell}, {Morales}, {Morgan},
  {Kudryavtseva}, {Oberoi}, {Ord}, {Pindor}, {Procopio}, {Prabu}, {Riding},
  {Roshi}, {Shankar}, {Srivani}, {Subrahmanyan}, {Tingay}, {Waterson},
  {Webster}, {Whitney}, {Williams}, \& {Williams}}]{Offringa2014}
{Offringa}, A.~R., {McKinley}, B., {Hurley-Walker}, N., {et~al.} 2014, \mnras,
  444, 606

\bibitem[{{Offringa} \& {Smirnov}(2017)}]{offringa2017}
{Offringa}, A.~R. \& {Smirnov}, O. 2017, \mnras, 471, 301

\bibitem[{{Peng} {et~al.}(2010){Peng}, {Lilly}, {Kova{\v c}}, {Bolzonella},
  {Pozzetti}, {Renzini}, {Zamorani}, {Ilbert}, {Knobel}, {Iovino}, {Maier},
  {Cucciati}, {Tasca}, {Carollo}, {Silverman}, {Kampczyk}, {de Ravel},
  {Sanders}, {Scoville}, {Contini}, {Mainieri}, {Scodeggio}, {Kneib}, {Le
  F{\`e}vre}, {Bardelli}, {Bongiorno}, {Caputi}, {Coppa}, {de la Torre},
  {Franzetti}, {Garilli}, {Lamareille}, {Le Borgne}, {Le Brun}, {Mignoli},
  {Perez Montero}, {Pello}, {Ricciardelli}, {Tanaka}, {Tresse}, {Vergani},
  {Welikala}, {Zucca}, {Oesch}, {Abbas}, {Barnes}, {Bordoloi}, {Bottini},
  {Cappi}, {Cassata}, {Cimatti}, {Fumana}, {Hasinger}, {Koekemoer},
  {Leauthaud}, {Maccagni}, {Marinoni}, {McCracken}, {Memeo}, {Meneux}, {Nair},
  {Porciani}, {Presotto}, \& {Scaramella}}]{Peng2010}
{Peng}, Y.-j., {Lilly}, S.~J., {Kova{\v c}}, K., {et~al.} 2010, \apj, 721, 193

\bibitem[{{Pillepich} {et~al.}(2019){Pillepich}, {Nelson}, {Springel},
  {Pakmor}, {Torrey}, {Weinberger}, {Vogelsberger}, {Marinacci}, {Genel}, {van
  der Wel}, \& {Hernquist}}]{Pillepich2019}
{Pillepich}, A., {Nelson}, D., {Springel}, V., {et~al.} 2019, \mnras, 490, 3196

\bibitem[{{Poggianti} {et~al.}(2019){Poggianti}, {Ignesti}, {Gitti}, {Wolter},
  {Brighenti}, {Biviano}, {George}, {Vulcani}, {Gullieuszik}, {Moretti},
  {Paladino}, {Bettoni}, {Franchetto}, {Jaff{\'e}}, {Radovich}, {Roediger},
  {Tomi{\v{c}}i{\'c}}, {Tonnesen}, {Bellhouse}, {Fritz}, \&
  {Omizzolo}}]{Poggianti2019xray}
{Poggianti}, B.~M., {Ignesti}, A., {Gitti}, M., {et~al.} 2019, \apj, 887, 155

\bibitem[{{Poggianti} {et~al.}(2017{\natexlab{a}}){Poggianti}, {Jaff{\'e}},
  {Moretti}, {Gullieuszik}, {Radovich}, {Tonnesen}, {Fritz}, {Bettoni},
  {Vulcani}, {Fasano}, {Bellhouse}, {Hau}, \& {Omizzolo}}]{Poggianti2017nat}
{Poggianti}, B.~M., {Jaff{\'e}}, Y.~L., {Moretti}, A., {et~al.}
  2017{\natexlab{a}}, \nat, 548, 304

\bibitem[{{Poggianti} {et~al.}(2017{\natexlab{b}}){Poggianti}, {Moretti},
  {Gullieuszik}, {Fritz}, {Jaff{\'e}}, {Bettoni}, {Fasano}, {Bellhouse}, {Hau},
  {Vulcani}, {Biviano}, {Omizzolo}, {Paccagnella}, {D'Onofrio}, {Cava},
  {Sheen}, {Couch}, \& {Owers}}]{Poggianti2017}
{Poggianti}, B.~M., {Moretti}, A., {Gullieuszik}, M., {et~al.}
  2017{\natexlab{b}}, \apj, 844, 48

\bibitem[{{Postman} \& {Geller}(1984)}]{Postman1984}
{Postman}, M. \& {Geller}, M.~J. 1984, \apj, 281, 95

\bibitem[{{Ramatsoku} {et~al.}(2020){Ramatsoku}, {Serra}, {Poggianti},
  {Moretti}, {Gullieuszik}, {Bettoni}, {Deb}, {Franchetto}, {van Gorkom},
  {Jaff{\'e}}, {Tonnesen}, {Verheijen}, {Vulcani}, {Andati}, {de Blok},
  {J{\'o}zsa}, {Kamphuis}, {Kleiner}, {Maccagni}, {Makhathini}, {Moln{\'a}r},
  {Ramaila}, {Smirnov}, \& {Thorat}}]{Ramatsoku2020}
{Ramatsoku}, M., {Serra}, P., {Poggianti}, B.~M., {et~al.} 2020, \aap, 640, A22

\bibitem[{{Ramatsoku} {et~al.}(2019){Ramatsoku}, {Serra}, {Poggianti},
  {Moretti}, {Gullieuszik}, {Bettoni}, {Deb}, {Fritz}, {van Gorkom},
  {Jaff{\'e}}, {Tonnesen}, {Verheijen}, {Vulcani}, {Hugo}, {J{\'o}zsa},
  {Maccagni}, {Makhathini}, {Ramaila}, {Smirnov}, \& {Thorat}}]{Ramatsoku2019}
{Ramatsoku}, M., {Serra}, P., {Poggianti}, B.~M., {et~al.} 2019, \mnras, 487,
  4580

\bibitem[{{Ramos-Mart{\'\i}nez} {et~al.}(2018){Ramos-Mart{\'\i}nez},
  {G{\'o}mez}, \& {P{\'e}rez-Villegas}}]{Ramos2018}
{Ramos-Mart{\'\i}nez}, M., {G{\'o}mez}, G.~C., \& {P{\'e}rez-Villegas}, {\'A}.
  2018, \mnras, 476, 3781

\bibitem[{{Randriamampandry} {et~al.}(2021){Randriamampandry}, {Wang}, \&
  {Mogotsi}}]{Randriamampandry2021}
{Randriamampandry}, T.~H., {Wang}, J., \& {Mogotsi}, K.~M. 2021, \apj, 916, 26

\bibitem[{{Rhee} {et~al.}(2017){Rhee}, {Smith}, {Choi}, {Yi}, {Jaff{\'e}},
  {Candlish}, \& {S{\'a}nchez-J{\'a}nssen}}]{Rhee2017}
{Rhee}, J., {Smith}, R., {Choi}, H., {et~al.} 2017, \apj, 843, 128

\bibitem[{{Roberts} {et~al.}(2021){Roberts}, {van Weeren}, {McGee}, {Botteon},
  {Ignesti}, \& {Rottgering}}]{Roberts2021}
{Roberts}, I.~D., {van Weeren}, R.~J., {McGee}, S.~L., {et~al.} 2021, \aap,
  652, A153

\bibitem[{{Roediger} \& {Br{\"u}ggen}(2008)}]{Roediger2008}
{Roediger}, E. \& {Br{\"u}ggen}, M. 2008, \mnras, 388, L89

\bibitem[{{Saintonge} {et~al.}(2016){Saintonge}, {Catinella}, {Cortese},
  {Genzel}, {Giovanelli}, {Haynes}, {Janowiecki}, {Kramer}, {Lutz},
  {Schiminovich}, {Tacconi}, {Wuyts}, \& {Accurso}}]{Saintonge2016}
{Saintonge}, A., {Catinella}, B., {Cortese}, L., {et~al.} 2016, \mnras, 462,
  1749

\bibitem[{{Saintonge} {et~al.}(2017){Saintonge}, {Catinella}, {Tacconi},
  {Kauffmann}, {Genzel}, {Cortese}, {Dav{\'e}}, {Fletcher},
  {Graci{\'a}-Carpio}, {Kramer}, {Heckman}, {Janowiecki}, {Lutz}, {Rosario},
  {Schiminovich}, {Schuster}, {Wang}, {Wuyts}, {Borthakur}, {Lamperti}, \&
  {Roberts-Borsani}}]{Saintonge2017}
{Saintonge}, A., {Catinella}, B., {Tacconi}, L.~J., {et~al.} 2017, \apjs, 233,
  22

\bibitem[{{Saraf} {et~al.}(2023){Saraf}, {Wong}, {Cortese}, \&
  {Koribalski}}]{Saraf2023}
{Saraf}, M., {Wong}, O.~I., {Cortese}, L., \& {Koribalski}, B.~S. 2023, \mnras,
  519, 4128

\bibitem[{{Serra} {et~al.}(2013){Serra}, {Koribalski}, {Duc}, {Oosterloo},
  {McDermid}, {Michel-Dansac}, {Emsellem}, {Cuillandre}, {Alatalo}, {Blitz},
  {Bois}, {Bournaud}, {Bureau}, {Cappellari}, {Crocker}, {Davies}, {Davis}, {de
  Zeeuw}, {Khochfar}, {Krajnovi{\'c}}, {Kuntschner}, {Lablanche}, {Morganti},
  {Naab}, {Sarzi}, {Scott}, {Weijmans}, \& {Young}}]{Serra2013}
{Serra}, P., {Koribalski}, B., {Duc}, P.-A., {et~al.} 2013, \mnras, 428, 370

\bibitem[{{Serra} {et~al.}(2023){Serra}, {Maccagni}, {Kleiner}, {Moln{\'a}r},
  {Ramatsoku}, {Loni}, {Loi}, {de Blok}, {Bryan}, {Dettmar}, {Frank}, {van
  Gorkom}, {Govoni}, {Iodice}, {J{\'o}zsa}, {Kamphuis}, {Kraan-Korteweg},
  {Loubser}, {Murgia}, {Oosterloo}, {Peletier}, {Pisano}, {Smith}, {Trager}, \&
  {Verheijen}}]{Serra2023}
{Serra}, P., {Maccagni}, F.~M., {Kleiner}, D., {et~al.} 2023, \aap, 673, A146

\bibitem[{{Serra} {et~al.}(2015){Serra}, {Westmeier}, {Giese}, {Jurek},
  {Fl{\"o}er}, {Popping}, {Winkel}, {van der Hulst}, {Meyer}, {Koribalski},
  {Staveley-Smith}, \& {Courtois}}]{Serra2015}
{Serra}, P., {Westmeier}, T., {Giese}, N., {et~al.} 2015, \mnras, 448, 1922

\bibitem[{{Sivanandam} {et~al.}(2010){Sivanandam}, {Rieke}, \&
  {Rieke}}]{Sivanandam2010}
{Sivanandam}, S., {Rieke}, M.~J., \& {Rieke}, G.~H. 2010, \apj, 717, 147

\bibitem[{{Skrutskie} {et~al.}(2006){Skrutskie}, {Cutri}, {Stiening},
  {Weinberg}, {Schneider}, {Carpenter}, {Beichman}, {Capps}, {Chester},
  {Elias}, {Huchra}, {Liebert}, {Lonsdale}, {Monet}, {Price}, {Seitzer},
  {Jarrett}, {Kirkpatrick}, {Gizis}, {Howard}, {Evans}, {Fowler}, {Fullmer},
  {Hurt}, {Light}, {Kopan}, {Marsh}, {McCallon}, {Tam}, {Van Dyk}, \&
  {Wheelock}}]{Skrutskie2006}
{Skrutskie}, M.~F., {Cutri}, R.~M., {Stiening}, R., {et~al.} 2006, \aj, 131,
  1163

\bibitem[{{Smith} {et~al.}(2010){Smith}, {Lucey}, {Hammer}, {Hornschemeier},
  {Carter}, {Hudson}, {Marzke}, {Mouhcine}, {Eftekharzadeh}, {James},
  {Khosroshahi}, {Kourkchi}, \& {Karick}}]{RSmith2010}
{Smith}, R.~J., {Lucey}, J.~R., {Hammer}, D., {et~al.} 2010, \mnras, 408, 1417

\bibitem[{{Sun} {et~al.}(2010){Sun}, {Donahue}, {Roediger}, {Nulsen}, {Voit},
  {Sarazin}, {Forman}, \& {Jones}}]{Sun2010}
{Sun}, M., {Donahue}, M., {Roediger}, E., {et~al.} 2010, \apj, 708, 946

\bibitem[{{Sun} {et~al.}(2007){Sun}, {Donahue}, \& {Voit}}]{Sun2007}
{Sun}, M., {Donahue}, M., \& {Voit}, G.~M. 2007, \apj, 671, 190

\bibitem[{{Sun} {et~al.}(2006){Sun}, {Jones}, {Forman}, {Nulsen}, {Donahue}, \&
  {Voit}}]{Sun2006}
{Sun}, M., {Jones}, C., {Forman}, W., {et~al.} 2006, \apjl, 637, L81

\bibitem[{Sun \& Vikhlinin(2005)}]{Vikhlinin2005}
Sun, M. \& Vikhlinin, A. 2005, The Astrophysical Journal, 621, 718

\bibitem[{{Verdugo} {et~al.}(2015){Verdugo}, {Combes}, {Dasyra}, {Salom{\'e}},
  \& {Braine}}]{Verdugo2015}
{Verdugo}, C., {Combes}, F., {Dasyra}, K., {Salom{\'e}}, P., \& {Braine}, J.
  2015, \aap, 582, A6

\bibitem[{{Vollmer} {et~al.}(2001){Vollmer}, {Cayatte}, {van Driel}, {Henning},
  {Kraan-Korteweg}, {Balkowski}, {Woudt}, \& {Duschl}}]{Vollmer2001}
{Vollmer}, B., {Cayatte}, V., {van Driel}, W., {et~al.} 2001, \aap, 369, 432

\bibitem[{{Waldron} {et~al.}(2023){Waldron}, {Sun}, {Luo}, {Laudari},
  {Chatzikos}, {Sivanandam}, {Kenney}, {J{\'a}chym}, {Voit}, {Donahue}, \&
  {Fossati}}]{Waldron2023}
{Waldron}, W., {Sun}, M., {Luo}, R., {et~al.} 2023, \mnras, 522, 173

\bibitem[{{Wang} {et~al.}(2016){Wang}, {Koribalski}, {Serra}, {van der Hulst},
  {Roychowdhury}, {Kamphuis}, \& {Chengalur}}]{Wang2016}
{Wang}, J., {Koribalski}, B.~S., {Serra}, P., {et~al.} 2016, \mnras, 460, 2143

\bibitem[{{Westmeier} {et~al.}(2021){Westmeier}, {Kitaeff}, {Pallot}, {Serra},
  {van der Hulst}, {Jurek}, {Elagali}, {For}, {Kleiner}, {Koribalski},
  {Lee-Waddell}, {Mould}, {Reynolds}, {Rhee}, \&
  {Staveley-Smith}}]{Westmeier2021}
{Westmeier}, T., {Kitaeff}, S., {Pallot}, D., {et~al.} 2021, \mnras, 506, 3962

\bibitem[{{Whitaker} {et~al.}(2012){Whitaker}, {van Dokkum}, {Brammer}, \&
  {Franx}}]{Whitaker2012}
{Whitaker}, K.~E., {van Dokkum}, P.~G., {Brammer}, G., \& {Franx}, M. 2012,
  \apjl, 754, L29

\bibitem[{{Woudt} {et~al.}(2008){Woudt}, {Kraan-Korteweg}, {Lucey}, {Fairall},
  \& {Moore}}]{Woudt2008}
{Woudt}, P.~A., {Kraan-Korteweg}, R.~C., {Lucey}, J., {Fairall}, A.~P., \&
  {Moore}, S.~A.~W. 2008, \mnras, 383, 445

\bibitem[{{Yagi} {et~al.}(2007){Yagi}, {Komiyama}, {Yoshida}, {Furusawa},
  {Kashikawa}, {Koyama}, \& {Okamura}}]{Yagi2007}
{Yagi}, M., {Komiyama}, Y., {Yoshida}, M., {et~al.} 2007, \apj, 660, 1209

\bibitem[{{Yagi} {et~al.}(2010){Yagi}, {Yoshida}, {Komiyama}, {Kashikawa},
  {Furusawa}, {Okamura}, {Graham}, {Miller}, {Carter}, {Mobasher}, \&
  {Jogee}}]{Yagi2010}
{Yagi}, M., {Yoshida}, M., {Komiyama}, Y., {et~al.} 2010, \aj, 140, 1814

\bibitem[{{Yoon} {et~al.}(2017){Yoon}, {Chung}, {Smith}, \&
  {Jaff{\'e}}}]{Yoon2017}
{Yoon}, H., {Chung}, A., {Smith}, R., \& {Jaff{\'e}}, Y.~L. 2017, \apj, 838, 81

\bibitem[{{Yoshida} {et~al.}(2002){Yoshida}, {Yagi}, {Okamura}, {Aoki},
  {Ohyama}, {Komiyama}, {Yasuda}, {Iye}, {Kashikawa}, {Doi}, {Furusawa},
  {Hamabe}, {Kimura}, {Miyazaki}, {Miyazaki}, {Nakata}, {Ouchi}, {Sekiguchi},
  {Shimasaku}, \& {Ohtani}}]{Yoshida2002}
{Yoshida}, M., {Yagi}, M., {Okamura}, S., {et~al.} 2002, \apj, 567, 118

\bibitem[{{Zabel} {et~al.}(2019){Zabel}, {Davis}, {Smith}, {Maddox}, {Bendo},
  {Peletier}, {Iodice}, {Venhola}, {Baes}, {Davies}, {de Looze}, {Gomez},
  {Grossi}, {Kenney}, {Serra}, {van de Voort}, {Vlahakis}, \&
  {Young}}]{Zabel2019}
{Zabel}, N., {Davis}, T.~A., {Smith}, M. W.~L., {et~al.} 2019, \mnras, 483,
  2251

\end{thebibliography}

\end{document}